\documentstyle[preprint,prb,aps]{revtex}
\begin{document}
\draft
\bibliographystyle{prsty}
\title{Elastic scattering and absorption of surface acoustic waves
by a quantum dot}
\author{Andreas Kn\"abchen and
	Yehoshua B.\ Levinson}
\address{Weizmann Institute of Science,
Department of Condensed Matter Physics,\\
76100 Rehovot, Israel}
\author{Ora Entin-Wohlman}
\address{School of Physics and Astronomy,
Raymond and Beverly Sackler Faculty of Exact Sciences,\\
Tel Aviv University, 69978 Tel Aviv, Israel}
\date{\today}
\maketitle%
\begin{abstract}
We study theoretically the piezoelectric
interaction of a surface acoustic wave (SAW)
with a two-dimensional electron gas confined to an isolated quantum dot. 
The electron motion in the dot is diffusive. The
electron-electron interaction is accounted for by
solving the screening problem in real space. Since the screening
in GaAs/Al${}_x$Ga${}_{1-x}$As
heterostructures is strong, an approximate inversion of the
dielectric function $\epsilon (\mbox{\boldmath $r$},\mbox{\boldmath $r$}')$ can be
utilized, providing a comprehensive qualitative picture of the screened SAW
potential and the charge redistribution in the dot. We calculate
the absorption and the scattering cross-sections for SAW's as a function
of the area of the dot, $A$, the sound wave vector, $q$, 
and the diffusion coefficient $D$
of the electrons. Approximate analytical expressions for
the cross-sections are derived for all cases where the quantities
$q^2A$ and $A\omega/D$ are much larger or smaller than unity; $\omega$
is the SAW frequency. 
Numerical results which
include the intermediate regimes and show the
sample-specific dependence of the cross-sections on the angles of
incidence and scattering of surface phonons are discussed.
The weak localization corrections to the cross-sections are found
and discussed as a function of a weak magnetic field, the frequency,
and the temperature.
Due to the absence of current-carrying contacts, the phase
coherence of the electron motion, and in turn the quantum corrections,
increase as the size of the dot shrinks. This shows that 
scattering and absorption of sound as noninvasive probes
may be advantageous in comparison to transport experiments
for the investigation of very small electronic systems.
\end{abstract}
% 7250 acoustoelectric effects
% 7335 mesoscopic systems
% 7215R quantum localization

\section{Introduction}\label{intro}

During the last decade, a number of theoretical papers
which address the application of ultrasound for the investigation of
quantum effects in disordered electronic systems has been published. 
Mainly quantum corrections to the sound absorption in infinite systems
have been studied. For instance,
the contribution of weak localization effects to the absorption coefficient
has been calculated in Refs.\
\cite{Houghton85,Kotliar85,Afonin86,Kirkpatrick86,Reizer89}.
Electron-electron interaction effects have been addressed
in Refs.\ \cite{Houghton85} and \cite{Kotliar85}. A particularly detailed
discussion of these effects, including both the diffusion and the
cooper channel terms, is given in Ref.\ \cite{Reizer89}.
The interaction of sound with electrons confined to
a finite mesoscopic system has only been studied with respect to the fluctuations
of the ultrasound absorption.\cite{Serota88,Hershfield91}
The main idea of these two works is that the ultrasound absorption is
a noninvasive  probe which can be used to investigate isolated metallic samples
[no leads attached].
In all these works, the calculations 
have essentially been done for the deformation potential interaction 
of bulk phonons with three-dimensional [3D] electron systems.
To ensure an efficient coupling to the 3D phonon wave, the dimensionality of the
electron system cannot be reduced, though this is necessary in order
to enhance the weak localization effects. To overcome the restrictions associated with
bulk phonons, we propose to
consider the interaction of surface acoustic waves
\cite{Farnell78,Mayer95} (SAW's)
with 2D electron systems.\cite{Rampton92} 
This interaction is very strong in GaAs/Al${}_x$Ga${}_{1-x}$As
heterostructures
where it is caused by the piezoelectric field accompanying the SAW.
Indeed, the SAW technique has been used successfully to investigate
both the integer
and the fractional quantum Hall 
regime.\cite{Wixforth86,Wixforth89,Schenstrom88,Willet90,Guillion91}
These experiments have shown, e.g., that the SAW technique is suited
to resolve very small spatial inhomogeneities in the
areal electron density which are not visible in dc magnetoresistance
measurements.\cite{Wixforth89,Guillion91}
Effects of electron heating due to the electric field accompanying
the SAW have been discussed in
Refs.\ \cite{Wixforth86} and \cite{Wixforth89}.
Though the absorption of SAW's in these experiments is used to study extended
electron systems, the SAW technique might be applied to mesoscopic systems as well.
In this case, the noninvasive character of such a measurement could prove advantageous.
In a very recent experiment,\cite{Shilton96} the direct acousto-electric
current induced by a SAW through a {\it single} quantum point
contact has been observed. The length of the quasi-one-dimensional channel
[which determines the size of the interaction region] was about
$0.5$ $\mu$m. 
%We note in passing that in a few papers
%results for the absorption of artficial 2D phonons by 2D electrons are given.
%However, the experimental feasibility of low-dimensional phonons is questionable.

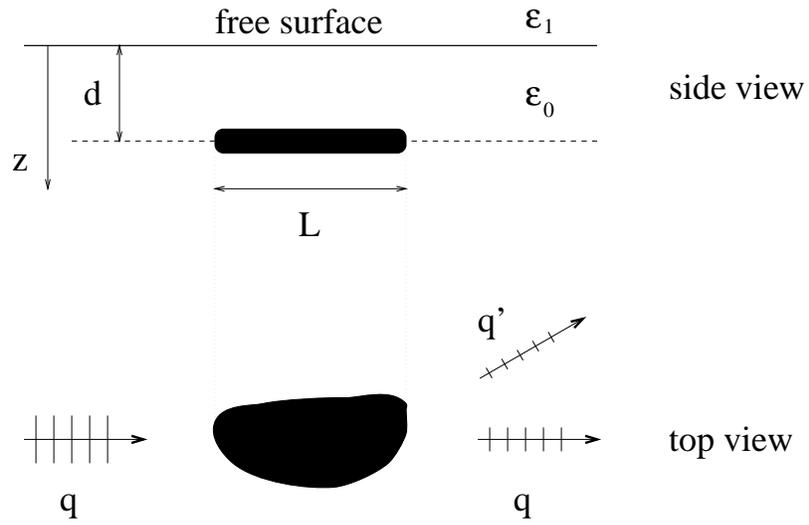
\begin{figure}
\caption{\label{model}
Schematic drawing of the experiment.
The quantum dot, shown in black, is of size $L$ and
is separated by a spacer layer of thickness $d$ from
the surface. The dielectric constants of the sample and the half-space
above it are denoted by $\epsilon_0$ and $\epsilon_1$, respectively.
The incoming and the transmitted waves
have the wave vector \protect\boldmath{$q$}, whereas 
\protect\boldmath{$q'$}
is the wave vector of the scattered wave.
}
\end{figure}

It is the main purpose of this paper to consider theoretically
some of the effects associated with a noninvasive probing of mesoscopic
2D electron systems by SAW's.
Specifically, we address
the scattering and absorption of SAW's due to the
electrons confined to an isolated quantum dot, see
Fig.\ \ref{model}. 
One of the main quantities to be calculated in this framework is the elastic
differential scattering cross-section $\eta_{sc}(\mbox{\boldmath $q'$}, \mbox{\boldmath $q$})$.
By definition, $\eta_{sc}d\varphi$ is the ratio of the sound intensity
flux scattered into a ``solid" angle $d\varphi$ around \mbox{\boldmath $q'$} and the flux
intensity $I$ of the incoming surface wave with wave vector
\mbox{\boldmath $q$}, $q=q'$.
Besides $\eta_{sc}$, we introduce the cross-section
$\eta_{abs}$ characterizing the phonon absorption.
[Though $\eta_{sc}$ and $\eta_{abs}$ have the
dimension of a length in two dimensions, we shall use the familiar term cross-section.]
$I\eta_{abs}$
gives the energy per unit time absorbed by the electrons in the dot from
the acoustic wave field. Hence, this quantity is directly associated with
electron heating.

We calculate the weak localization corrections to both
$\eta_{sc}$ and $\eta_{abs}$. Since the sample is isolated,
the phase coherence is not reduced by leads which are necessarily
attached to the dot in an electron
transport measurement. This in turn affects the magnitude of the weak localization
corrections and their dependence on the size of the dot.
Though weak localization effects contribute only correction terms to the
classical cross-sections, their dependence
on weak magnetic fields and the phase coherence time (i.e.\ the temperature)
can be used to detect them. Their particular dependence on the frequency
is superimposed on that of the classical components of $\eta_{sc}$ and $\eta_{abs}$
and might therefore be difficult to resolve.

The screening of the electron-phonon coupling 
arises from the electrons confined to the dot and is not a negligible effect.
We account for the screening in the linear response approximation, where
the change of the electron density arising from the external perturbation is
proportional to the magnitude of the perturbation.
This approach is justified by the small SAW intensities used in experiments.
Since we consider
a system without translational invariance, the equations for
the screened potential, the charge redistribution, etc.\ have to be
formulated in real space. Consequently, screening cannot be taken into account
by simply multiplying the unscreened potential by a dielectric function
$\epsilon(\omega,\mbox{\boldmath $q$})$ but involves
the inversion of the dielectric function (or matrix) 
$\epsilon(\mbox{\boldmath $r$},\mbox{\boldmath $r$}')$. 
To do this accurately, i.e.\ to account for the shape of the dot
and the direction of the incoming SAW, we have performed numerical
calculations. Analytically, 
one can exploit the fact that the screening is strong.
In particular, for wavelengths $2\pi/q$ which are larger than the size $L$
of the dot, a series expansion of $\epsilon^{-1}$ in terms of the
small parameter $a_B/L$ can be utilized,  where $a_B$ is the effective Bohr
radius.
This provides a rather complete qualitative understanding of the relations
between the bare and the screened SAW potential and the charge redistribution
in the dot.

In the calculation of the cross-sections $\eta_{sc}$ and $\eta_{abs}$
we mainly focus on the 
cases where $qL$ is of the order of or smaller than unity.
The quantum dot will be considered in the diffusive limit,
i.e.\ the size $L$ of the dot is large compared to the elastic mean free path $l$.
In addition, $l$ has to be small compared to the wavelength $2\pi/q$
of the SAW, $ql \ll 1$. This relation
guarantees $\omega \tau \ll 1$, because the velocity of sound, $s=\omega/q$, is much
smaller than the Fermi velocity, $v_F=l/\tau$. From an experimental point of view,
these conditions are satisfied in a dot of size $L \simeq 1$ $\mu$m,
patterened in an electron gas with a low
mobility ($\mu \simeq 10^4$ cm$^2/$Vs) corresponding
to $l \simeq 100$ nm. So, except for the shortest SAW's used in recent 
experiments,\cite{Wixforth89,Schenstrom88,Willet90,Guillion91,Rampton92,Shilton96}
$ql$ is indeed small.

This paper is organized in the following way.
In Sec.\ II, we summarize the main equations for the scattering and
the absorption cross-sections, the bare SAW potential arising from the piezoelectric
coupling, and the dielectric function.
The cross-sections $\eta_{sc}$ and $\eta_{abs}$ and, 
within the linear screening approach, the dielectric function 
$\epsilon(\mbox{\boldmath $r$},\mbox{\boldmath $r$}')$
are essentially determined by the density-density correlator
$\Pi_\omega(\mbox{\boldmath $r$},\mbox{\boldmath $r$}')$. This quantity is specified
in Sec.\ III for the case of a diffusive system, where it comprises 
besides the classical term weak localization corrections. Based on these results,
we discuss $\epsilon(\mbox{\boldmath $r$},\mbox{\boldmath $r$}')$ and its appropriate matrix
representation in Sec.\ \ref{screen}. An approximate inversion of $\epsilon$
is carried out analytically in the strong screening regime. This yields 
the screened potential in the dot in terms of the bare
SAW field. Combining these results with the equations for
$\Pi_\omega(\mbox{\boldmath $r$},\mbox{\boldmath $r$}')$, 
we evaluate the cross-sections in the limiting
cases  $qL \ll 1$ and $qL \gg 1$ in Sec.\ \ref{cross}.
The weak localization corrections to  $\eta_{sc}$ and $\eta_{abs}$ are related to
the cooperon ${\cal C}_\omega (\mbox{\boldmath $r$},\mbox{\boldmath $r$})$. 
Its equation is solved in Sec.\
\ref{cooper}. Special emphasis is put on the dependence of 
${\cal C}_\omega (\mbox{\boldmath $r$},\mbox{\boldmath $r$})$ on the magnetic field. Results of
a numerical computation of the scattering and the absorption
cross-sections are presented
in Sec.\ \ref{numerics}. Conclusions are given in the last section.

\section{Basic equations}

The interaction between the SAW and the electrons in the quantum dot
gives rise to finite probabilities for the absorption and the scattering
of phonons. A Golden Rule calculation can be used to obtain the
corresponding cross-sections. The amplitude for the absorption of a phonon results from
a first-order
process between the (phonon) states
$|\mbox{\boldmath $q$}\rangle$ and $|0 \rangle$, see Fig.\ \ref{ephver};
$\mbox{\boldmath $q$}$ is the 2D
phonon wave vector. Scattering is a second-order
process involving the two intermediate states 
$|\mbox{\boldmath $q$},\mbox{\boldmath $q$}'\rangle$
and $|0,0\rangle $ with two or no phonons,
depending on whether the emission of the second phonon occurs before
or after the absorption of the incoming one.
The absorption and (elastic) scattering
cross-sections have the form 
(a factor of 2 accounting for the spin degeneracy is included)
\begin{equation}\label{etaab}
\eta_{abs}(\mbox{\boldmath $q$}) = -\frac{4 {\cal L}^2 }{s\hbar} \Im [
\Pi_{\omega} (\mbox{\boldmath $q$},\mbox{\boldmath $q$})]  ,
\end{equation}%
and
\begin{equation}\label{etasc}
\eta_{sc}(\mbox{\boldmath $q'$}, 
\mbox{\boldmath $q$}) = \frac{q {\cal L}^4 }{\pi s^2 \hbar^2}
| \Pi_{\omega}( \mbox{\boldmath $q$}', \mbox{\boldmath $q$}) |^2  ,
\end{equation}%
where $s$ is the velocity of surface sound and
\begin{equation}\label{piomqq}
\Pi_{\omega} (\mbox{\boldmath $q$},\mbox{\boldmath $q$}')\equiv
\int d^3\mbox{\boldmath $R$} \int d^3\mbox{\boldmath $R$}' \,
V^*_{\mbox{\boldmath $q$}} (\mbox{\boldmath $R$}) 
\Pi_\omega(\mbox{\boldmath $R$},\mbox{\boldmath $R$}') 
V_{\mbox{\boldmath $q$}'} (\mbox{\boldmath $R$}')  .
\end{equation}%
Here, $\mbox{\boldmath $R$}$ is a 3D real space vector
and $V_{\mbox{\boldmath $q$}}$ denotes the {\it
screened} potential associated with one surface phonon with
wave vector $\mbox{\boldmath $q$}$ in the normalization area ${\cal L}^2$.
The quantity ${\cal L}$ does not enter the final results
since it is canceled by corresponding terms originating form 
the SAW potential, see Eq.\ (\ref{poten}) below.
The retarded density-density
correlator $\Pi_\omega (\mbox{\boldmath $R$}, \mbox{\boldmath $R'$})$ 
of the electrons in the dot is
defined by\cite{Fetter71}
\begin{equation}\label{corgen}
\Pi_\omega (\mbox{\boldmath $R$}, \mbox{\boldmath $R'$}) =
-(i/\hbar)
\int_0^\infty dt \, e^{i\omega t} \left\langle [\rho(\mbox{\boldmath $R$},t), 
\rho(\mbox{\boldmath $R'$},0)]
\right\rangle  ,
\end{equation}%
where $\rho(\mbox{\boldmath $R$},t)$ is the electron density operator 
for one spin component and
$\omega=qs$ is the SAW frequency.

Equation (\ref{piomqq}) can be simplified by making use of the fact that
the thickness of the 2D electron gas (2DEG) is much smaller than
the penetration depth of the SAW into the interior of the sample.
This allows one to neglect the finite extend
of the 2DEG in the $z$-direction, replacing $\rho(\mbox{\boldmath $R$},t)$ by
$\delta(z-d) \rho(\mbox{\boldmath $r$},t)$, 
where $d$ is the distance between the 2DEG and the surface
of the sample and $\rho(\mbox{\boldmath $r$},t)$ is the areal density of 2D electrons.
[We have $\mbox{\boldmath $R$}=(\mbox{\boldmath $r$},z)$ where $\mbox{\boldmath $r$}$ 
is a vector
in the plane of the 2DEG and $z$ is the co-ordinate
perpendicular to it; see Fig.\ \ref{model}.]
Substituting this replacement into Eq.\ (\ref{corgen})
yields
\begin{equation}\label{denpi3}
\Pi_\omega(\mbox{\boldmath $R$},\mbox{\boldmath $R$}')= \delta(z-d) \delta(z'-d) 
\Pi_\omega(\mbox{\boldmath $r$},\mbox{\boldmath $r$}') 
 ,
\end{equation}%
where $\Pi_\omega(\mbox{\boldmath $r$},\mbox{\boldmath $r$}')$ 
is the remaining 2D density-density correlator.
Particle number conservation
can be expressed in terms of $\Pi_\omega(\mbox{\boldmath $r$},\mbox{\boldmath $r$}')$ in the form
\begin{equation}\label{parconv}
\int d^2\mbox{\boldmath $r$} \, \Pi_\omega(\mbox{\boldmath $r$},\mbox{\boldmath $r$}')=
\int d^2\mbox{\boldmath $r$}' \, \Pi_\omega(\mbox{\boldmath $r$},\mbox{\boldmath $r$}')=0  .
\end{equation}%
Substituting Eq.\ (\ref{denpi3}) into Eq.\ (\ref{piomqq}), we obtain
\begin{equation}\label{piqrr}
\Pi_{\omega} (\mbox{\boldmath $q$},\mbox{\boldmath $q$}')\equiv
\int d^2\mbox{\boldmath $r$} \int d^2\mbox{\boldmath $r$}' \,
V^*_{\mbox{\boldmath $q$}} (\mbox{\boldmath $r$},z=d) 
\Pi_\omega(\mbox{\boldmath $r$},\mbox{\boldmath $r$}') 
V_{\mbox{\boldmath $q$}'} (\mbox{\boldmath $r$}', z'=d)  .
\end{equation}%
The integrations run over the area
$A$ of the dot.

The bare potential $V^{ph}$ created
by the SAW in the plane of the 2DEG
can be represented in the form\cite{Knabchen96}
\begin{equation}\label{poten}
V^{ph}_{\mbox{\boldmath $q$}}(\mbox{\boldmath $r$}, z=d)= \frac{1}{{\cal L}} 
\gamma_{\mbox{\boldmath $q$}}
e^{i\mbox{\boldmath $q$}\mbox{\boldmath $r$}}  .
\end{equation}%
For GaAs/Al${}_x$Ga${}_{1-x}$As
heterostructures and the
range of wavelengths used in SAW experiments,
the piezoelectric electron-phonon interaction is dominant.
We may thus identify $\gamma_{\mbox{\boldmath $q$}}$ with the piezoelectric vertex 
$\gamma_{\mbox{\boldmath $q$}}^{PA}$, neglecting
the deformation potential coupling.
In addition, since $qd$ is usually much smaller
than unity, the dependence of $\gamma_{\mbox{\boldmath $q$}}^{PA}$ on $d$ can be disregarded.
[For $qd \sim 1$, $\gamma_{\mbox{\boldmath $q$}}$ depends non-monotonously on the
parameter $qd$, see the discussion in Ref.\ \cite{Simon96}.]
Then, we have
\begin{equation}\label{gamma}
\gamma_{\mbox{\boldmath $q$}} = \gamma_{\mbox{\boldmath $q$}}^{PA} =
(\hbar/\rho s a_{PA})^{1/2} \beta e \hat{q}_x \hat{q}_y
= 3.7 \hat{q}_x \hat{q}_y 10^{-10} \,\,\, {\rm eV cm}  ,
\end{equation}%
where $\rho$ is the mass density of the lattice, $e$ is the electron charge, and 
$a_{PA}$ represents a numerical factor
which can be expressed in terms of the elastic constants
of the lattice, cf.\ Ref.\ \cite{Knabchen96}.
Equation (\ref{gamma})
is valid for a GaAs-type crystal with the SAW propagating along the (100) plane
and electrically free\cite{Farnell78} boundary conditions
for the piezoelectric potential at the surface. In this case all (non-zero)
piezoelectric moduli are equal to $\beta$. $\hat{q}_x$ (and, similarly, $\hat{q}_y$)
is the component of $\hat{\mbox{\boldmath $q$}}$ in the direction of the lattice axis $x$
on the surface.
The numerical value given on the right-hand-side of Eq.\ (\ref{gamma}) applies to
GaAs/Al${}_x$Ga${}_{1-x}$As
heterostructures.

The potential
$V^{ph}$ associated with the SAW acts on the electrons in the dot and leads to their
redistribution. This creates a potential
$V^{ch}$ which adds to $V^{ph}$.
The resulting total potential
\begin{equation}\label{vtot}
V=V^{ch}+V^{ph}
\end{equation}%
is the relevant quantity which determines the absorption and the scattering
of surface phonons by the quantum dot; see Eq.\ (\ref{piomqq}).
The calculation of the total potential
$V$ and the corresponding charge redistribution
$\delta \rho(\mbox{\boldmath $r$})$ has to be done self-consistently. 
Although the electron distribution has been restricted to a plane, the
electrostatic problem is still a three-dimensional one.
Bearing in mind that the quantum dot is embedded
in a semiconductor with dielectric constant $\epsilon_\circ$,
we can write the following equations\cite{Fetter71}
\begin{equation}\label{deltarho}
\delta \rho(\mbox{\boldmath $r$}) =
2  \int d^2\mbox{\boldmath $r$}' \, \Pi_\omega(\mbox{\boldmath $r$},\mbox{\boldmath $r$}') 
V(\mbox{\boldmath $r$}', z'=d)  ,
\end{equation}%
\begin{equation}\label{laplace}
\nabla^2 V^{ch}(\mbox{\boldmath $R$})  = 
-\frac{4\pi e^2}{\epsilon_\circ} \delta(z-d) \delta \rho(\mbox{\boldmath $r$})  ,
\end{equation}%
where the factor of 2 is due to spin degeneracy and
it is understood that all potentials and $\delta \rho(\mbox{\boldmath $r$})$
refer to the $\omega$-component in the corresponding
Fourier expansions.
Note that
Eqs.\ (\ref{vtot})--(\ref{laplace}) reduce for a translational
invariant system to the well-known Random Phase Approximation for the dielectric function.

The solution of Poisson's equation (\ref{laplace})
can be expressed in terms of the corresponding Green's function
\begin{equation}\label{green}
\nabla^2 G(\mbox{\boldmath $R$},\mbox{\boldmath $R$}') = -4\pi 
\delta(\mbox{\boldmath $R$}-\mbox{\boldmath $R$}')
\end{equation}%
which has to satisfy the boundary conditions at
the interface between the sample and the halfspace (dielectric
constant $\epsilon_1$) above it. (The
SAW potential $V^{ph}$ satisfies the boundary conditions so that
the total potential $V$ meets all requirements provided that $V^{ch}$ does.)
Addressing the case where both $\mbox{\boldmath $R$}$ and $\mbox{\boldmath $R$}'$ lie in
the plane of the dot ($z=z'=d$), we have\cite{Landau84}
\begin{equation}\label{greenex}
G(\mbox{\boldmath $R$},\mbox{\boldmath $R$}')=G(\mbox{\boldmath $r$}-\mbox{\boldmath $r$}') 
= \frac{1}{|\mbox{\boldmath $r$}-\mbox{\boldmath $r$}'|}
+ \frac{\epsilon_\circ-\epsilon_1}{\epsilon_\circ+\epsilon_1} \,
\frac{1}{\sqrt{|\mbox{\boldmath $r$}-\mbox{\boldmath $r$}'|^2+(2d)^2}} .
\end{equation}%
The Green's function $G$ can be combined with Eqs.\
(\ref{vtot})--(\ref{laplace}) to relate the total potential directly 
to the SAW field
\begin{equation}\label{vvph}
\int d^2\mbox{\boldmath $r$}' \, \epsilon (\mbox{\boldmath $r$},\mbox{\boldmath $r$}') 
V(\mbox{\boldmath $r$}', z'=d)
= V^{ph} (\mbox{\boldmath $r$}, z=d) .
\end{equation}
The kernel of this integral equation is the dielectric function
\begin{equation}\label{eps}
\epsilon (\mbox{\boldmath $r$},\mbox{\boldmath $r$}')=
\delta (\mbox{\boldmath $r$}-\mbox{\boldmath $r$}')
- 2 \frac{e^2}{\epsilon_\circ} \int d^2\mbox{\boldmath $r$}'' \,
G(\mbox{\boldmath $r$}- \mbox{\boldmath $r$}'') \Pi_\omega( \mbox{\boldmath $r$}'', 
\mbox{\boldmath $r$}')   .
\end{equation}%
Using Eq.\ (\ref{parconv}), one can see that
$\int d^2\mbox{\boldmath $r$}' \, \epsilon (\mbox{\boldmath $r$},\mbox{\boldmath $r$}')=1$. 
This means that
a potential which is spatially constant within the dot is not screened,
cf.\ Eq.\ (\ref{vvph}). 

\section{Density-density correlator for a diffusive system}\label{dense}

In the diffusive regime, the density-density correlator\cite{Altshuler92}
has the form
\begin{equation}\label{corr}
\Pi_\omega ( \mbox{\boldmath $r$}, \mbox{\boldmath $r$}' )=
- \nu \left[ \delta(\mbox{\boldmath $r$}-\mbox{\boldmath $r$}') +
i \omega {\cal D}_\omega (\mbox{\boldmath $r$},\mbox{\boldmath $r$}')
\right]  ,
\end{equation}%
where ${\cal D}_\omega$ is the diffusion propagator and
$\nu$ is the (2D) density of states for one spin projection.
This result is valid for small frequencies $\omega \tau \ll 1$, 
small wave vectors $q l \ll 1$,
and low temperatures $\omega, T \ll \epsilon_F$; $\tau$ and $l$ denote
the elastic mean free time and mean free path, respectively, and
$\epsilon_F$ is the Fermi energy. $q$ describes the spatial modulation of
an external potential, the response to which can be expressed in terms
of $\Pi_\omega$, Eq.\ (\ref{corr}). In our case, $q$ is the wave number 
of the SAW's. 
Neglecting weak localization corrections, the
diffusion propagator 
satisfies in real space the equation\cite{Altshuler92}
\begin{equation}\label{difprop}
[ -i \omega - D \nabla^2 ] {\cal D}_\omega (\mbox{\boldmath $r$}, \mbox{\boldmath $r'$} )
= \delta(\mbox{\boldmath $r$}-\mbox{\boldmath $r'$})
\qquad \nabla_n {\cal D}_\omega|_b =0  ,
\end{equation}%
where $D=l^2/2\tau$ is the 2D diffusion coefficient and
the outer normal component of a vector [here the gradient]
with respect to the boundary of the quantum dot is denoted by a subscript $n$.
The boundary condition follows from the requirement that there is no flow
of electrons through the boundary of the system. This is in contrast
to a system coupled to leads, where
the particle density is fixed in the contact regions, i.e.\ ${\cal D}|_c=0$.

The diffusion propagator, Eq.\ (\ref{difprop}),
can be expressed in terms of its eigenfunctions as follows
\begin{equation}\label{difpro2}
{\cal D}_\omega(\mbox{\boldmath $r$},\mbox{\boldmath $r'$}) =
\sum\limits_{m=0}^\infty \frac{\psi_m(\mbox{\boldmath $r$}) 
\psi_m(\mbox{\boldmath $r$}')}{-i\omega + D\lambda_m}  .
\end{equation}%
The diffusion modes are defined by
\begin{equation}\label{difmodes}
[\nabla^2 + \lambda_m ] \psi_m(\mbox{\boldmath $r$})=0, \qquad \nabla_n \psi_m|_b=0  .
\end{equation}%
They are orthogonal to each other and are normalized, 
$\int d^2\mbox{\boldmath $r$} \, \psi_m \psi_n=\delta_{m,n}$.
It is a peculiar feature of an isolated quantum dot
that there exists a zeroth eigenfunction $\psi_0 = 1/\sqrt{A}$,
$A$ being the area of the dot.
The corresponding eigenvalue $\lambda_0=0$ is well
separated from the remaining sequence of eigenvalues $\lambda_m \sim A^{-1}$.
The zeroth mode determines the behavior of the diffusion propagator
in the case of a ``small" dot, $A\omega/D \ll 1$. In this regime,
the particle is able to diffuse through the whole system within one period
of the external potential. Boundary effects are crucial and we obtain
from Eq.\ (\ref{difpro2}) ${\cal D}_\omega(\mbox{\boldmath $r$},\mbox{\boldmath $r'$})
\simeq (-i\omega A)^{-1}$.
In the opposite, ``big-dot" case, $A\omega/D \gg 1$,
the particle diffuses only over a distance
$\sqrt{D/\omega } \ll \sqrt{A}$
before the external potential is reversed.
Thus, the diffusion process is bulk-like. Disregarding all
boundary effects, Eq.\ (\ref{difpro2}) reduces in this case
to the translational invariant form 
${\cal D}_\omega(\mbox{\boldmath $r$}-\mbox{\boldmath $r'$})$, 
corresponding to an infinitely extended system.
In the intermediate regime, $A\omega/D \simeq 1$, the diffusion propagator
exhibits sample-specific properties.

Weak localization effects yield a correction 
term $\delta D$ 
to the classical diffusion coefficient $D$ which basically describes
the slowing down of the diffusion processes due to enhanced backscattering.
\cite{Altshuler92,Vollhardt80} Generally,
$\delta D$ may depend on the frequency $\omega$, 
the electron phase coherence time $\tau_\phi$,
a weak magnetic field $B$, and other physical parameters. 
In addition to this, the weak localization correction to
$D$ acquires in a finite system a spatial dependence.
To account for a spatially varying diffusion coefficient in
Eq.\ (\ref{difprop}) for the diffusion propagator, we use the replacement\cite{Vollhardt80}
\begin{equation}\label{replac}
D\nabla^2 \longrightarrow 
\nabla ( D+\delta D(\mbox{\boldmath $r$})) \nabla  .
\end{equation}%
where all other variables of $\delta D$ are suppressed. 
This replacement guarantees particle number conservation.
Since we are only interested in the first order corrections due to
$\delta D(\mbox{\boldmath $r$})$, we write the diffusion propagator in the
from ${\cal D}_\omega+ \delta {\cal D}_\omega$. Substituting this ansatz
in the modified Eq.\ (\ref{difprop}) yields
\begin{equation}\label{deldif}
\delta {\cal D}_\omega (\mbox{\boldmath $r$}, \mbox{\boldmath $r'$})=
\int d^2\mbox{\boldmath $r$}'' \, 
{\cal D}_\omega (\mbox{\boldmath $r$}, \mbox{\boldmath $r$}'') 
\{ \nabla'' \delta D(\mbox{\boldmath $r$}'') \nabla'' \}
{\cal D}_\omega (\mbox{\boldmath $r$}'',\mbox{\boldmath $r$}')  .
\end{equation}%

Neglecting spin scattering,
the weak localization correction to
the diffusion coefficient can be expressed in
terms of the cooperon\cite{Vollhardt80,Altshuler92} ${\cal C}$ as follows
\begin{equation}\label{diffr}
\delta D(\mbox{\boldmath $r$}) =
-\frac{D}{\pi\hbar \nu} {\cal C}_\omega (\mbox{\boldmath $r$},\mbox{\boldmath $r$})  .
\end{equation}%
In real space, 
the cooperon obeys the equation\cite{Altshuler92}
\begin{eqnarray}\label{coop}
[ -i \omega + \tau_\phi^{-1} + D(i\nabla + (2e/c\hbar)\mbox{\boldmath $A$} 
(\mbox{\boldmath $r$}) )^2 ]
{\cal C}_\omega (\mbox{\boldmath $r$},\mbox{\boldmath $r'$}) & = & 
\delta(\mbox{\boldmath $r$}-\mbox{\boldmath $r'$}) \\
(i\nabla_n + (2e/c\hbar)\mbox{\boldmath $A$}_n (\mbox{\boldmath $r$}) ) 
{\cal C}_\omega|_b & = & 0  . \nonumber
\end{eqnarray}%
The influence of a (weak)
magnetic field $\mbox{\boldmath $B$}$ is described by the vector potential 
$\mbox{\boldmath $A$}(\mbox{\boldmath $r$})$.
The field $\mbox{\boldmath $B$}$ is oriented perpendicularly to the plane of the 2DEG.
The boundary condition in Eq.\ (\ref{coop})
ensures that there is no
flow of ``coherence" (${\cal C}$) through the
boundary of an isolated system. In contrast, the phase randomization
provided by a massive contact is described by ${\cal C}|_c=0$.
Here, we do not proceed with the evaluation of the cooperon;
this will be done in Sec.\ \ref{cooper}. For the rest
of this and the next two sections it will be sufficient to bear in mind that
$\delta D(\mbox{\boldmath $r$})$ is a well-defined quantity which can be
calculated according to Eq.\ (\ref{diffr}).

Let us now return to the density-density correlator. Substituting
Eqs.\ (\ref{difpro2}) and (\ref{deldif}) into Eq.\
(\ref{corr}) yields $\Pi_\omega$ in terms of the diffusion modes in the form
\begin{equation}\label{dcgen}
\Pi_\omega(\mbox{\boldmath $r$},\mbox{\boldmath $r$}')=
-\nu \sum_{m,n= 1}^\infty
\beta_{mn} \psi_m(\mbox{\boldmath $r$}) \psi_n(\mbox{\boldmath $r$}')  ,
\end{equation}%
where
\begin{equation}\label{begen}
\beta_{mn}=
\beta_m \delta_{m,n} + \delta\beta_{mn}
\end{equation}%
is decomposed into the classical term
\begin{equation}\label{becl}
\beta_m=
\frac{D\lambda_m}{-i\omega + D\lambda_m}
\end{equation}%
and the weak localization contribution
\begin{equation}\label{bewl}
\delta\beta_{mn}=
\frac{-i\omega}{(-i\omega+D\lambda_m)(-i\omega+D\lambda_n)}
\int d^2\mbox{\boldmath $r$} \, \delta D(\mbox{\boldmath $r$}) \nabla \psi_m(\mbox{\boldmath $r$})
\nabla \psi_n(\mbox{\boldmath $r$})  .
\end{equation}%
The sums over modes in Eq.\ (\ref{dcgen}) start from $m,n=1$ since
$\beta_{m0}= \beta_{0m}=0$ for $m=0,1,\ldots$. It can easily be seen
that this is a consequence of the structure of the diffusion propagator
[Eqs.\ (\ref{difprop}) and (\ref{replac})] and holds true even for
the case where $\delta D(\mbox{\boldmath $r$})$ is treated exactly (i.e.\ not only to first
order). The restriction of the summations means that, while
the zeroth mode contributes to the diffusion propagator, it does not
influence the density-density correlator.
The latter fulfills Eqs.\ (\ref{parconv}) because $\int d^2\mbox{\boldmath $r$} \, \psi_m=0$
for $m\ge 1$.

\section{Screening}\label{screen}

In order to apply Eq.\ (\ref{vvph}), the relation between the bare
SAW field and the screened potential $V$, to the diffusive dot under
consideration, we
consider its representation in terms 
of the diffusion modes defined in Eq.\ (\ref{difmodes}).
The matrix elements of the density-density correlator are given
in Eq.\ (\ref{dcgen}), while those of the Green's function $G$ [Eq.\
(\ref{greenex})] can be written as
\begin{equation}\label{gmn}
G_{mn}= \int d^2\mbox{\boldmath $r$}' \int d^2\mbox{\boldmath $r$} \,
\psi_m(\mbox{\boldmath $r$}) G(\mbox{\boldmath $r$}-\mbox{\boldmath $r$}') 
\psi_n(\mbox{\boldmath $r$}')  .
\end{equation}%
For the potential $V$ (and, similarly, for $V^{ph}$), we introduce the expansion 
\begin{equation}\label{vinvers}
V(\mbox{\boldmath $r$}, z=d)=\sum\limits_{n=0}^\infty V_n \psi_n(\mbox{\boldmath $r$})  ,
\qquad 
V_n=\int d^2\mbox{\boldmath $r$} \, \psi_n(\mbox{\boldmath $r$}) V(\mbox{\boldmath $r$}, z=d)  .
\end{equation}%
Using these definitions, the complete set of equations which follows from Eq.\
(\ref{vvph}) can be written in the form
\begin{equation}\label{voph}
V^{ph}_0 = V_0 + 2 \frac{e^2 \nu}{\epsilon_\circ}
\sum_{n,l \ge 1} G_{0l} \beta_{ln} V_n  ,
\end{equation}%
\begin{equation}\label{vmph}
V^{ph}_m= V_m +2 \frac{e^2 \nu}{\epsilon_\circ}
\sum_{n,l \ge 1} G_{ml} \beta_{ln} V_n \, , \quad m\ge 1.
\end{equation}%
These equations have to be solved with respect to $\{V_n\}$. Not all of these
quantities are coupled to each other. For example, as emphasized by Eqs.\
(\ref{voph}) and (\ref{vmph}), the $\{ V_n \}$, $n\ge 1$, form a closed
system of equations. Its solution can be substituted into Eq.\ (\ref{voph})
determining the element $V_0$. This property of the screening
equations results from the fact that, due to charge conservation,
$\beta_{l0}=0$, cf.\ the discussion after Eq.\ (\ref{bewl}).

The formal solution of Eq.\ (\ref{vmph}) can be given 
in terms of an inverse dielectric matrix [$m\ge 1$]
\begin{equation}\label{epsin}
V_m= \sum_{n\ge 1} (\epsilon^{-1})_{mn} V_n^{ph}  .
\end{equation}%
A precise calculation of the elements $(\epsilon^{-1})_{mn}$ 
has to be done numerically. This is described in Sec.\ \ref{numerics}.
Here, we shall exploit the following facts. 
First, the SAW potential is slowly varying on the scale of the dot.
Indeed, for usual sound frequencies,
$V^{ph}$ oscillates once or a few times across the dot. Consequently,
only the first few elements $V^{ph}_n$ are significantly
different from zero. With respect to the matrices $G_{mn}$, $\beta_{mn}$,
and $(\epsilon^{-1})_{mn}$, we may also concentrate on the indices 
$m$ and $n$ which are of order unity. 
Secondly,
the screening in experimentally relevant samples is strong.
To see this, we estimate the magnitude of the different
terms in Eqs.\ (\ref{voph}) and (\ref{vmph}).
Using $\nu=m^*/2\pi\hbar^2$, the prefactor $2 e^2 \nu/\epsilon_\circ$
can be written in the form $1/\pi a_B$, where
$a_B=\epsilon_\circ/e^2 m^*$ is the effective Bohr radius
of the lattice and $m^*$ is the effective electron mass.
Since $a_B =10.6$ nm for GaAs, the Bohr radius represents the smallest
length scale in the system.
The matrix element $G_{mn}$ is of order 
$1/\sqrt{\lambda_m} \simeq L/m$ for $m\approx n $, $L\equiv \sqrt{A}$ being
the size of the dot,
and it
decreases sharply for $m$ or $n$ much larger 
than unity and very different from each other.
Hence, we have $(2 e^2 \nu/\epsilon_\circ) G_{mn} \simeq L/a_B$ 
for the relevant $m$ and $n $ of order unity.
We therefore expect $V_m$ to be of order $(a_B/L) V_m^{ph} \ll V_m^{ph}$.

An approximate inversion of Eq.\ (\ref{vmph}) providing the leading terms
in an expansion with respect to $a_B/L$ can be accomplished by introducing
the inverse matrix $(G^{-1})_{ml}$ to the reduced matrix $G_{ml}$ with indices
$m$ and $l$ equal to or larger than unity.
Similarly, we define
$(\beta^{-1})_{ml}$. Multiplying Eq.\ (\ref{vmph}) with $\beta^{-1}G^{-1}$
yields for the inverse dielectric operator [Eq.\
(\ref{epsin})]
\begin{eqnarray}\label{epsinex}
(\epsilon^{-1})_{mn} & = &
\frac{\pi a_B}{L} \sum_{l\ge 1}
(\beta^{-1})_{ml} (\tilde G^{-1})_{ln} + {\cal O} ( a_B^2/L^2)  , 
\end{eqnarray}%
where $\tilde G\equiv G/ L$ is a dimensionless Green's function depending
only on the shape of the dot.

Substituting Eqs.\ (\ref{epsinex}) and (\ref{epsin}) into Eq.\
(\ref{voph}), we obtain for the $\mbox{\boldmath $r$}$-independent part of the total
potential
\begin{equation}\label{vzero}
V_0 = V_0^{ph} - \sum_{n,l \ge 1} G_{0l} (G^{-1})_{ln} V_n^{ph} + {\cal O} (a_B/L)  .
\end{equation}%
This equation confirms explicitly the conclusion following from 
the general Eq.\ (\ref{eps}), namely that the spatially uniform part
of an external potential, here $V^{ph}_0$, contributes unscreened to $V_0$.
Moreover, since the product $G G^{-1}$ is of order unity,
it shows that also the spatially varying components $V_n^{ph}$
contribute effectively unscreened to $V_0$.
Combining Eqs.\ (\ref{epsinex}) and (\ref{vzero}),
we find for the total potential 
\begin{equation}\label{vcomp}
V(\mbox{\boldmath $r$}) \approx \frac{1}{\sqrt{A}} V_0 +
\frac{\pi a_B}{L} \sum_{m,n,l \ge 1} \psi_m(\mbox{\boldmath $r$}) (\beta^{-1})_{mn}
(\tilde G^{-1})_{nl} V^{ph}_l
 .
\end{equation}%
This  result shows that, in contrast to an open or an infinitely extended system, the
case of strong screening in an isolated dot is characterized
by small variations of the total potential (the second term) existing on top of a large but
spatially constant background (the first term). 
The background term is of the same
order as $V^{ph}$.
That is, an isolated quantum dot
is not able to completely screen out an external potential.
This behavior is based on particle conservation, for the charge on
the dot can only be redistributed to some extent but cannot be increased
or reduced via electrons flowing to or coming from the
leads.
Absorption and scattering of phonons are associated with the spatially
varying component of the total potential which carries the 
factor $a_B/L \ll 1$.
The screening of the SAW potential by the electrons in the quantum dot
is thus an effect which reduces considerably the magnitude of
the scattering and the absorption cross-sections.

In concluding this section let us
consider the charge redistribution $\delta \rho(\mbox{\boldmath $r$})$.
Substituting Eqs.\ (\ref{dcgen}) and (\ref{vcomp}) into
Eq.\ (\ref{deltarho}) yields
\begin{eqnarray}\label{deltaex}
\delta \rho(\mbox{\boldmath $r$}) &=&
-2  \nu \delta(z-d) \sum\limits_{m,n\ge 1} \psi_m(\mbox{\boldmath $r$}) \beta_{mn} V_n \\
& = & -2  \nu \delta(z-d) \frac{\pi a_B}{L}
\sum_{m,n \ge 1} \psi_m(\mbox{\boldmath $r$}) (\tilde G^{-1})_{mn} V_n^{ph}
+ {\cal O} (a_B^2/L^2)  .\nonumber
\end{eqnarray}%
That is, even in the strong screening case, where $a_B$ is very small
compared to other length scales, $\delta \rho(\mbox{\boldmath $r$})$ is determined
by the distribution of the external potential within the whole dot. Indeed,
the $V_n^{ph}$ couple via non-diagonal elements of $\tilde G^{-1}$
to other modes $m$.
In this sense, screening in an isolated dot is strongly non-local.

\section{Scattering and absorption cross-sections}\label{cross}

In this section we study the absorption and the scattering cross-sections,
Eqs.\ (\ref{etaab}) and (\ref{etasc}), in the limiting cases $qL \gg 1$
and $qL \ll 1$. We focus on the dependences of the
cross-sections on $q$, $D$, and the area $A$ of the dot, and on the qualitative influence
of the weak localization corrections. The substantiation of these analytical
results by numerical calculations, addressing also
the angular dependence of the cross-sections, their sensitivity to the
shape of the dot, etc., will be deferred until Sec.\ \ref{numerics}.

\subsection{The case $q\protect L\protect\gg 1$}\label{qalarge}

This regime resembles the case of an infinitely extended system.
One may therefore use the usual $\mbox{\boldmath $q$}$-space representation
for the density-density correlator, the dielectric function, etc.
This leads to the simple relation 
$V(\mbox{\boldmath $r$})=V^{ph}_{\mbox{\boldmath $q$}}
(\mbox{\boldmath $r$})/\epsilon(\omega,\mbox{\boldmath $q$})$
between the total potential and the SAW field. The dielectric function 
\begin{equation}\label{emninf}
\epsilon(\omega,\mbox{\boldmath $q$})=
1+\frac{2\pi e^2}{\epsilon_0 q} 
\left[ 1+\frac{1-\epsilon_1/\epsilon_0}{1+\epsilon_1/\epsilon_0} e^{-qd}
\right] 2 \nu \beta(\omega,\mbox{\boldmath $q$}) \approx
1+2 \frac{\beta(\omega,\mbox{\boldmath $q$})}{a_B q}
\end{equation}%
is derived form Eq.\ (\ref{eps}). Here,
\begin{equation}\label{betaoq}
\beta(\omega,\mbox{\boldmath $q$})= \frac{ (D+\delta D) q^2}{ -i\omega +(D+ \delta D)q^2}
\end{equation}%
is (except for the factor $-\nu$)
the Fourier representation of the density-density correlator, Eq.\ (\ref{denpi3}),
which replaces the expression $\beta_{mn}$ [Eq.\ (\ref{dcgen})] valid in the
diffusion mode representation. To obtain Eq.\ (\ref{betaoq}),
the weak localization correction $\delta D(\mbox{\boldmath $r$})$
to the diffusion coefficient [Eq.\ (\ref{diffr})]
has been replaced
by some average value $\delta D=$ const, and, consequently, is not $\mbox{\boldmath $r$}$-dependent.
Using $qd\ll 1$ and $\epsilon_1= 1\ll
\epsilon_\circ$, the dielectric function simplifies to the result given
on the right-hand-side of Eq.\ (\ref{emninf}).
The latter condition corresponds
to vacuum above the surface of the semiconductor. The effective
dielectric constant in the vicinity of the surface (distance smaller than
$1/q$) is then given by $(\epsilon_\circ+\epsilon_1)/2 \approx \epsilon_\circ/2$.

Substituting the bare SAW potential [Eq.\ (\ref{poten})] and
the dielectric function $\epsilon(\omega,\mbox{\boldmath $q$})$ [Eq.\
(\ref{emninf})] in Eq.\ (\ref{piqrr}) for the quantity $\Pi_\omega(\mbox{\boldmath $q$},
\mbox{\boldmath $q$}')$,
we obtain for the absorption cross-section [Eq.\ (\ref{etaab})]
\begin{equation}\label{etaabinfp}
\eta_{abs}(\mbox{\boldmath $q$})/A = 
\frac{4 \nu}{s \hbar} |\gamma_{\mbox{\boldmath $q$}}|^2 \, 
\frac{ \Im [\beta(\omega,\mbox{\boldmath $q$})] }{ |\epsilon(\omega,\mbox{\boldmath $q$})|^2 } 
\equiv \Gamma_{\mbox{\boldmath $q$}}
 . 
\end{equation}%
The attenuation coefficient $\Gamma_{\mbox{\boldmath $q$}}$ is the relevant
quantity for an extended system as it describes
the decrease of the intensity of a SAW traveling a distance $x$ 
along the surface as exp$(-\Gamma_{\mbox{\boldmath $q$}}x)$.
Note that $\eta_{abs}(\mbox{\boldmath $q$})$
does not possess the meaning of the total energy absorbed
from the SAW once $\Gamma_{\mbox{\boldmath $q$}} L$ becomes larger than unity.
Neglecting weak localization corrections, the attenuation coefficient
given in Eq.\ (\ref{etaabinfp}) coincides with the result following
from the well-known treatment of sound absorption due
to the piezoelectric interaction
[see, e.g., Refs.\ \cite{Wixforth89} and \cite{Simon96}].
In the case of strong screening, Eq.\ (\ref{etaabinfp}) can be written in the form
\begin{equation}\label{etaabinf}
\Gamma_{\mbox{\boldmath $q$}} = 
\frac{ \nu}{s \hbar} |\gamma_{\mbox{\boldmath $q$}}|^2 (a_B)^2 \frac{\omega}{D}
\left(1-\frac{\Re[\delta D]}{D} \right)
=
\frac{1}{2} K_{\text{eff}}^2 q \frac{\sigma_m}{\sigma}
\left(1-\frac{\Re[\delta \sigma]}{\sigma} \right)  ,
\end{equation}%
where  the right-hand-side uses the ``standard" notation, i.e.\ 
$\Gamma_{\mbox{\boldmath $q$}}$ is given in terms of the 2D conductivity $\sigma$,
the conductivity $\sigma_m\equiv \epsilon_0 s/4\pi$, and the
effective electromechanical coupling coefficient $K_{\text{eff}}^2=
|\gamma_{\mbox{\boldmath $q$}}|^2 \epsilon_0 /2 \pi s e^2 \hbar$.
Equation (\ref{etaabinf}) does not only reproduce the well-known
classical result for the absorption coefficient but gives also its dependence
on the weak localization effects expressed in terms of $\delta D$ or the
weak localization correction $\delta \sigma$ of the conductivity.
Since $\Re[ \delta D] \sim \Re[ \delta \sigma ] <0$,
they enhance the absorption. The primary reason for this enhancement is
the reduced screening caused by the slowing down of the diffusion processes.
The enhancement factor $1-\Re[\delta D]/D$ has been
found in previous work
\cite{Houghton85,Afonin86,Kirkpatrick86,Reizer89}
on the absorption of bulk sound in a 3D electron
system or of (hypothetical) 2D phonons by a 2DEG,
independent of whether the piezoelectric or the deformation potential
electron-phonon interaction has been  studied. A different result has been obtained
in Ref.\ \cite{Kotliar85}. The reason for this,
as discussed in Ref.\ \cite{Kirkpatrick86}, is the insufficient number
of diagrams incorporated in that calculation.

Equation (\ref{etasc}) for the scattering cross-section can be treated similarly.
In the limit $q L \gg 1$, $\eta_{sc}$ 
has a dominating forward-scattering component
\begin{equation}\label{etascinf}
\eta_{sc}(\mbox{\boldmath $q$}',\mbox{\boldmath $q$}) 
\sim \delta(\mbox{\boldmath $q$}'-\mbox{\boldmath $q$})  .
\end{equation}%
This property results from the momentum conservation
in a translational invariant system.

\subsection{The case $qL \ll 1$}

Here, we exploit the diffusion mode representation introduced in the 
previous two sections. Substituting the density-density correlator,
Eq.\ (\ref{dcgen}), and the total potential in the strong screening limit, Eq.\
(\ref{vcomp}), into Eq.\ (\ref{piqrr}) yields
\begin{equation}\label{piq2}
\Pi_\omega(\mbox{\boldmath $q$},\mbox{\boldmath $q$}') =
- \nu \frac{(\pi a_B)^2}{A} 
\sum_{m,n,k,l \ge 1}
(\beta^{-1})^*_{mn} (\tilde G^{-1})_{nk} (V^{ph}_k)^*
(\tilde G^{-1})_{ml} V^{ph}_l .
\end{equation}%
Expanding the bare SAW potential [Eq.\ (\ref{poten})] in a series with respect to
$|\mbox{\boldmath $q$}\mbox{\boldmath $r$}| \ll 1$, we find
\begin{equation}\label{vphex}
\sum_{l\ge 1} (\tilde G^{-1})_{ml} V^{ph}_l 
=
\frac{1}{{\cal L}} \gamma_{\mbox{\boldmath $q$}} i q A a_m(\hat{\mbox{\boldmath $q$}})  ,
\end{equation}%
where
\begin{equation}\label{amq}
a_m(\hat{\mbox{\boldmath $q$}}) \equiv
\sum_{n\ge 1} (\tilde G^{-1})_{mn} 
\int \frac{d^2\mbox{\boldmath $r$}}{A} \, \hat{\mbox{\boldmath $q$}} \mbox{\boldmath $r$}
\psi_n(\mbox{\boldmath $r$})  .
\end{equation}%
The dimensionless integral in this equation is of order unity  for small $n$'s,
and it decreases as $n$ increases. $a_m(\hat{\mbox{\boldmath $q$}})$ is expected to have the same
properties. Introducing result (\ref{vphex}) in Eq.\ (\ref{piq2}),
we obtain 
\begin{equation}\label{piq3}
\Pi_\omega(\mbox{\boldmath $q$},\mbox{\boldmath $q$}') =
- \nu (\pi a_B \gamma_{\mbox{\boldmath $q$}} q)^2 A {\cal L}^{-2}
\sum_{m,n \ge 1}
(\beta^{-1})^*_{mn} a_m(\hat{\mbox{\boldmath $q$}}) a_n(\hat{\mbox{\boldmath $q$}'})  .
\end{equation}%
Up to first order in the weak localization corrections,
the inverse of $\beta_{mn}$, Eq.\ (\ref{begen}), is given by
\begin{equation}\label{begeninv}
(\beta^{-1})_{mn} =
\beta_m^{-1} \delta_{m,n} - \frac{\delta \beta_{mn}}{\beta_m \beta_n} 
\end{equation}%
where $\delta \beta_{mn}$ is defined in Eq.\ (\ref{bewl}).

We are now in the position to use Eq.\ (\ref{piq3}) in the evaluation
of the absorption and the scattering cross-sections.
Substituting Eq.\ (\ref{piq3}) into Eq.\ (\ref{etaab})
yields
\begin{eqnarray}\label{etaabqap}
\eta_{abs}(\mbox{\boldmath $q$}) & = &
\frac{4 \nu}{s \hbar} |\gamma_{\mbox{\boldmath $q$}}|^2 (\pi a_B)^2
q^2A^2\frac{\omega}{D} \\
&&
\times \sum_{m,n\ge 1} \frac{1}{A \lambda_m}
\left( \delta_{m,n} - \frac{1}{\lambda_n} \int d^2\mbox{\boldmath $r$} \,
\nabla \psi_m(\mbox{\boldmath $r$}) 
\nabla \psi_n(\mbox{\boldmath $r$}) \frac{\Re[\delta D(\mbox{\boldmath $r$})]}{D} \right)
a_m(\hat{\mbox{\boldmath $q$}}) a_n(\hat{\mbox{\boldmath $q$}})  . \nonumber
\end{eqnarray}%
Note that $A \lambda_m$ is a dimensionless quantity independent of $A$, cf.\
Eq.\ (\ref{difmodes}).
A significant simplification of this equation is achieved when $\delta D(\mbox{\boldmath $r$})$
does not vary in space. This is not always the case, of course. We believe, however,
that, qualitatively, the influence of the weak localization
effects is described by an average quantity $\delta D=$ const, which will be defined
in Sec.\ \ref{cooper}. Replacing $\delta D(\mbox{\boldmath $r$})$ by $\delta D$ and using
the diffusion mode equation (\ref{difmodes}), Eq.\ (\ref{etaabqap})
reduces to
\begin{equation}\label{etaabqa}
\eta_{abs}(\mbox{\boldmath $q$}) =
\frac{4 \nu}{\hbar} |\gamma_{\mbox{\boldmath $q$}}|^2 (\pi a_B)^2
A^2\frac{q^3}{D}
\left( 1- \frac{\Re[\delta D] }{D} \right) 
\left\{ \sum_{m\ge 1} \frac{a^2_m(\hat{\mbox{\boldmath $q$}})}{A \lambda_m} \right\}  .
\end{equation}%
The dependence on the shape of the dot and the direction of the SAW is comprised
in the quantity in braces. The dependences on all other parameters
is completely described by its prefactor.
As in the case of an infinite
system, Eq.\ (\ref{etaabinf}), the weak localization corrections enhance the absorption. This
can be again understood as a result of the reduced screening of the SAW potential.
Comparing Eqs.\ (\ref{etaabinf}) and (\ref{etaabqa}) we see that
$\eta_{abs}$ is smaller by a factor $(q^2A)$ in the case $qL \ll 1$.
That is, a small system absorbs per unit area much less than an extended one.
This can be attributed to the fact that the electrons in an extended system
can move over the whole period $1/q$ of the piezoelectric field, while
a small system restricts this motion by its size.

We shall now evaluate the scattering cross-section $\eta_{sc}$.
Substituting Eq.\ (\ref{piq3}) into Eq.\ (\ref{etasc}) yields
\begin{eqnarray}\label{etascp}
\eta_{sc}(\mbox{\boldmath $q'$}, \mbox{\boldmath $q$}) & = &
\frac{q \nu^2}{ \pi s^2 \hbar^2 }
(\pi a_B q )^4 A^2 \\
&& \times \sum_{m,n,k,l\ge 1}
(\beta^{-1})_{mn}^* (\beta^{-1})_{kl} 
a_m(\hat{\mbox{\boldmath $q$}}') a_n(\hat{\mbox{\boldmath $q$}})
 a_k (\hat{\mbox{\boldmath $q$}}') a_l(\hat{\mbox{\boldmath $q$}})
 . \nonumber
\end{eqnarray}%
Here, we should replace the matrix elements of $\beta^{-1}$
by the explicit expressions given in Eq.\ (\ref{begeninv}) and consider the classical
and weak localization contributions separately. As far as the latter are
concerned, we have to calculate the sum over four $a$-terms with the $\beta^{-1}\beta^{-1}$-part
replaced by
\begin{displaymath}
-2 \Re \left[ \frac{1}{\beta_m^*} \frac{\delta \beta_{kl}}{\beta_k\beta_l} 
\right]  .
\end{displaymath}%
This can be rewritten as
\begin{equation}\label{help3}
-2 \frac{\omega}{D^2 \lambda_k\lambda_l} \int d^2\mbox{\boldmath $r$} \,
\nabla \psi_k(\mbox{\boldmath $r$}) \nabla \psi_l(\mbox{\boldmath $r$})
\{ \Im[\delta D(\mbox{\boldmath $r$})] + 
\frac{\omega}{D \lambda_m} \Re[\delta D(\mbox{\boldmath $r$})] \}  ,
\end{equation}%
showing that the relevant quantity is either $\Re[\delta D(\mbox{\boldmath $r$})]$
or $\Im[\delta D(\mbox{\boldmath $r$})]$ depending on whether the diffusion time
through the dot, $A/D$, is
large or small compared to the period of the SAW potential.

The replacement of $\delta D(\mbox{\boldmath $r$})$ by $\delta D=$const leads to a considerable
simplification of expression (\ref{help3}).
Using this approximation to evaluate Eq.\ (\ref{etascp}), we obtain explicit
estimates in the limits of a small and a big dot:
\begin{eqnarray}\label{etascqa}
\eta_{sc}(\mbox{\boldmath $q'$}, \mbox{\boldmath $q$}) & \approx &
\frac{\nu^2}{ \pi s^2 \hbar^2 } |\gamma_{\mbox{\boldmath $q$}}|^2
|\gamma_{\mbox{\boldmath $q$}'}|^2 (\pi a_B)^4 q^5 A^2 \\
&& 
\times
\sum_{m\ge 1} a_m(\hat{\mbox{\boldmath $q$}}') a_m(\hat{\mbox{\boldmath $q$}})
\sum_{n\ge 1} a_n(\hat{\mbox{\boldmath $q$}}') a_n(\hat{\mbox{\boldmath $q$}})
\{ 1-2\frac{\omega }{D\lambda_n} \frac{\Im[\delta D]}{D} \} \quad
\omega A/D \ll 1 \nonumber\\
&&
\times
\left( \sum_{m\ge 1} \frac{a_m(\hat{\mbox{\boldmath $q$}}') 
a_m(\hat{\mbox{\boldmath $q$}})}{A\lambda_m}
\right)^2
\left( \frac{\omega A}{D} \right)^2
\{ 1- 2\frac{\Re[\delta D]}{D} \} \qquad \qquad \omega A/D \gg 1 \nonumber
\end{eqnarray}%
The classical contribution and the prefactor of the weak localization
corrections to the scattering cross-section
depend strongly on the parameter $A\omega/D$.
For a big dot, $A\omega/D>1$, the scattering of SAW's
rises faster than $A^2$ with increasing area $A$ of the dot
because the diffusion processes are too slow to screen
long wavelength density variations as effectively as short 
wavelength ones.
Indeed, the enhancement factor 
$\omega A/D$ results from the classical part of $(\beta^{-1})_{mn}$,
$\beta_m^{-1}=1-i\omega/D\lambda_m$ [Eq.\ (\ref{bewl})],
which enters Eq.\ (\ref{etascp}) via the inverse of the dielectric matrix.
In the small dot case, $\beta_m^{-1} \simeq 1$ for all $m$ starting from unity,
and, hence, there is no enhancement factor in the second line of Eq.\ (\ref{etascqa}).
While the quantum corrections
contribute to $\eta_{sc}$ according to their relative magnitude for
big dots, they acquire a small prefactor once $\omega A/D$ becomes
smaller than unity. Moreover, since their imaginary part is the relevant
quantity in that case, they are small for $\omega < \tau_\phi^{-1}$,
cf.\ Eq.\ (\ref{smalldot}) below. This is contrary to the
dependence of weak localization corrections 
on frequency and phase coherence time as far as the real part of the conductivity
is concerned.

\section{Weak localization corrections}\label{cooper}

The absorption and the scattering cross-sections depend on weak
localization corrections via $\delta D(\mbox{\boldmath $r$})$ which in turn
is directly related to the cooperon ${\cal C}_\omega(\mbox{\boldmath $r$},\mbox{\boldmath $r$})$;
see Eq.\ (\ref{diffr}). In the first part of this section,
we evaluate the cooperon equation (\ref{coop}). Special attention
is devoted to the magnetic field dependence. In the second part,
we discuss the approximation of $\delta D(\mbox{\boldmath $r$})$ by a
spatially independent quantity $\delta D$.

\subsection{Magnetic field dependence}\label{coopermag}

In comparison with the diffusion propagator, Eq.\ (\ref{difpro2}),
the cooperon, Eq.\ (\ref{coop}), depends on the two additional length scales
$l_\phi = \sqrt{D \tau_\phi}$ and $l_B= \sqrt{c\hbar/2e B}$.
Due to the sensitivity of quantum corrections to weak magnetic fields,
it is sufficient to account perturbatively for the $B$-dependent terms
in Eq.\ (\ref{coop}).
To this end, we expand the cooperon
in a power series with respect to $\mbox{\boldmath $A$}(\mbox{\boldmath $r$})$:
\begin{equation}\label{coex}
{\cal C}_\omega (\mbox{\boldmath $r$},\mbox{\boldmath $r$}')=
{\cal C}^{(0)}_\omega  (\mbox{\boldmath $r$},\mbox{\boldmath $r$}') +
{\cal C}^{(1)}_\omega  (\mbox{\boldmath $r$},\mbox{\boldmath $r$}') +
{\cal C}^{(2)}_\omega (\mbox{\boldmath $r$},\mbox{\boldmath $r$}') + \ldots  ,
\end{equation}%
where ${\cal C}^{(m)}_\omega  \sim \mbox{\boldmath $A$}^m$.
This expansion can be terminated with a negligible error at the first
nonvanishing correction to
${\cal C}^{(0)}_\omega (\mbox{\boldmath $r$},\mbox{\boldmath $r$})$ if
$L/l_B \ll 1$ and $|-i \omega + \tau_\phi^{-1}| \gg DL^2/l_B^4$.
Using experimental values [which are summarized in Sec.\ \ref{numerics}],
we have $\omega \lesssim \tau_\phi^{-1}$, i.e.\ the second inequality can be written
as $Ll_\phi/l_B^2 \ll 1$. Since $l_\phi \lesssim L$ in the cases of practical interest,
the first condition $L/l_B \ll 1$ determines the range of applicability of the
perturbative treatment.
Substituting Eq.\ (\ref{coex}) in Eq.\ (\ref{coop})
yields
\begin{eqnarray}\label{set}
[ -i \omega + \tau_\phi^{-1} - D\nabla^2 ]
{\cal C}_\omega^{(0)} (\mbox{\boldmath $r$},\mbox{\boldmath $r'$})&  =&  
\delta(\mbox{\boldmath $r$}-\mbox{\boldmath $r'$}) 
 , \nonumber\\
\protect[-i \omega + \tau_\phi^{-1} - D\nabla^2 ]
{\cal C}_\omega^{(1)} (\mbox{\boldmath $r$},\mbox{\boldmath $r'$}) &  =& 
-2iD(2e/c\hbar)\mbox{\boldmath $A$}(\mbox{\boldmath $r$}) 
\nabla {\cal C}_\omega^{(0)} (\mbox{\boldmath $r$},\mbox{\boldmath $r'$})
 , \\
\protect[-i \omega + \tau_\phi^{-1} - D\nabla^2 ]
{\cal C}_\omega^{(2)} (\mbox{\boldmath $r$},\mbox{\boldmath $r'$}) &  =&
-2iD(2e/c\hbar)\mbox{\boldmath $A$}(\mbox{\boldmath $r$}) \nabla 
{\cal C}_\omega^{(1)} (\mbox{\boldmath $r$},\mbox{\boldmath $r'$})
-D((2e/c\hbar)\mbox{\boldmath $A$}(\mbox{\boldmath $r$}))^2 
{\cal C}_\omega^{(0)} (\mbox{\boldmath $r$},\mbox{\boldmath $r'$})
 ,
\nonumber
\end{eqnarray}%
where we have used the gauge $\nabla \mbox{\boldmath $A$}(\mbox{\boldmath $r$})=0$.
The boundary conditions are given by
\begin{equation}\label{bccoop}
\nabla_n {\cal C}_\omega^{(0)}|_b  =  0  ,
\qquad \nabla_n {\cal C}_\omega^{(m+1)}|_b  =  
i(2e/c\hbar)\mbox{\boldmath $A$}_n{\cal C}_\omega^{(m)} |_b  , 
\, \mbox{\rm for} \,\, m =0,1,\ldots\, .
\end{equation}%
The equation for ${\cal C}^{(0)}_\omega$ can be solved using the diffusion
modes defined in Eq.\ (\ref{difmodes}) 
\begin{equation}\label{czero}
{\cal C}_\omega^{(0)} (\mbox{\boldmath $r$}, \mbox{\boldmath $r$}')=
\sum\limits_{m=0}^{\lambda_m \lesssim l^{-2}}
\frac{\psi_m(\mbox{\boldmath $r$}) \psi_m(\mbox{\boldmath $r$}')}{-i\omega
+ D\lambda_m + \tau_\phi^{-1}}  .
\end{equation}%
The cut-off on the summation is because
the diffusion approximation is valid on scales larger than the
mean free path.
The relevant frequency scale
is given by $|-i\omega +\tau_\phi^{-1}|$.

Since ${\cal C}^{(0)}_\omega $, Eq.\ (\ref{czero}),
is the Green's function for the differential operator
of all three equations (\ref{set}), it enables us to write down formal solutions
for ${\cal C}^{(1)}_\omega $ and ${\cal C}^{(2)}_\omega $ as well.
The result for ${\cal C}^{(1)}_\omega $ reads
\begin{eqnarray}\label{c1int}
{\cal C}_\omega^{(1)} (\mbox{\boldmath $r$},\mbox{\boldmath $r'$}) & = &
D \oint d\mbox{\boldmath $S$}_1 \, 
[i(2e/c\hbar)\mbox{\boldmath $A$}_n(\mbox{\boldmath $r$}_1)
{\cal C}_\omega^{(0)} (\mbox{\boldmath $r$}_1,\mbox{\boldmath $r'$})]
{\cal C}_\omega^{(0)} (\mbox{\boldmath $r$}_1,\mbox{\boldmath $r$}) \nonumber
\\
&& +
\int d^2\mbox{\boldmath $r$}_1 \, 
[-2iD(2e/c\hbar)\mbox{\boldmath $A$}(\mbox{\boldmath $r$}_1)) 
\nabla_1 {\cal C}_\omega^{(0)} (\mbox{\boldmath $r$}_1,\mbox{\boldmath $r'$}) ]
{\cal C}_\omega^{(0)} (\mbox{\boldmath $r$}_1,\mbox{\boldmath $r$})  .
\end{eqnarray}%
The first term accounts for the source term in the differential
equation, whereas the second one represents an integral over
the boundary of the dot and includes the inhomogeneity
of the boundary condition (\ref{bccoop}).
Equation (\ref{c1int}) can be rewritten in the symmetric form
\begin{equation}\label{c1ints}
{\cal C}_\omega^{(1)} (\mbox{\boldmath $r$},\mbox{\boldmath $r'$}) = 
\int d^2\mbox{\boldmath $r$}_1 \,
i (2e/c\hbar)\mbox{\boldmath $A$}(\mbox{\boldmath $r$}_1) \nabla_1
[ {\cal C}_\omega^{(0)} (\stackrel{\downarrow}{\mbox{\boldmath $r$}}_1,\mbox{\boldmath $r$})
{\cal C}_\omega^{(0)} (\mbox{\boldmath $r$}_1,\mbox{\boldmath $r'$})-
{\cal C}_\omega^{(0)} (\stackrel{\downarrow}{\mbox{\boldmath $r$}}_1,\mbox{\boldmath $r'$})
{\cal C}_\omega^{(0)} (\mbox{\boldmath $r$}_1,\mbox{\boldmath $r$})]  ,
\end{equation}%
where the arrow indicates the argument upon which the derivative
acts.
The last equation shows that ${\cal C}_\omega^{(1)}  
(\mbox{\boldmath $r$},\mbox{\boldmath $r$})=0$,
i.e.\ the quantum corrections to the diffusion coefficient
(\ref{diffr}) do not depend linearly on the magnetic field.
The expression for ${\cal C}^{(2)}_\omega$ has the same structure as Eq.\
(\ref{c1int}), one has just to substitute the corresponding source terms
given in Eqs.\ (\ref{set}). That is,
\begin{eqnarray}\label{c2int}
&&
{\cal C}_\omega^{(2)} (\mbox{\boldmath $r$},\mbox{\boldmath $r'$})  = 
D \oint d\mbox{\boldmath $S$}_1 \,
[i(2e/c\hbar)\mbox{\boldmath $A$}_n(\mbox{\boldmath $r$}_1)
{\cal C}_\omega^{(1)} (\mbox{\boldmath $r$}_1,\mbox{\boldmath $r'$})]
{\cal C}_\omega^{(0)} (\mbox{\boldmath $r$}_1,\mbox{\boldmath $r$})
\\
&&
+ \int d^2\mbox{\boldmath $r$}_1 \,
[-2iD(2e/c\hbar)\mbox{\boldmath $A$}(\mbox{\boldmath $r$}_1)) 
\nabla_1 {\cal C}_\omega^{(1)} (\mbox{\boldmath $r$}_1,\mbox{\boldmath $r'$}) 
-D((2e/c\hbar)\mbox{\boldmath $A$}(\mbox{\boldmath $r$}_1))^2 
{\cal C}_\omega^{(0)} (\mbox{\boldmath $r$}_1,\mbox{\boldmath $r$}')]
{\cal C}_\omega^{(0)} (\mbox{\boldmath $r$}_1,\mbox{\boldmath $r$})  . \nonumber
\end{eqnarray}%
This equation and Eq.\ (\ref{czero}) determine the weak
localization correction to the diffusion coefficient
in the form $\delta D(\mbox{\boldmath $r$})=-(D/\pi \hbar \nu) ( {\cal C}_\omega^{(0)}
(\mbox{\boldmath $r$},\mbox{\boldmath $r$}) + 
{\cal C}_\omega^{(2)} (\mbox{\boldmath $r$},\mbox{\boldmath $r$}))$.
Since ${\cal C}^{(2)}_\omega \sim B^2$, the physical quantities
calculated in the previous section are invariant with respect to a reversal
of the direction of the magnetic field.

\subsection{The average quantity $\protect \delta D$}\label{cooperdd}

The spatially constant quantity
$\delta D$ which was used in Sec.\ \ref{cross} 
is introduced by
\begin{equation}\label{deltad}
\delta D(\mbox{\boldmath $r$}) \longrightarrow \delta D \equiv
-\frac{D}{\pi \hbar \nu A} \int d^2\mbox{\boldmath $r$} \,
{\cal C}_\omega(\mbox{\boldmath $r$},\mbox{\boldmath $r$})  .
\end{equation}%
This approximation captures the essential features of the problem
and becomes exact in two limiting cases. For small frequencies
and large phase coherence times,
$|-i\omega +\tau_\phi^{-1}| \ll D/A$,
the cooperon is determined by the zeroth diffusion mode
$\psi_0$, and hence ${\cal C}_\omega(\mbox{\boldmath $r$},\mbox{\boldmath $r$}) \approx $const.
In the opposite case,
the bulk-like diffusion process guarantees that at least
${\cal C}^{(0)}_\omega(\mbox{\boldmath $r$},\mbox{\boldmath $r$})$ is spatially uniform
except near the boundaries.
In the intermediate regime, $|-i\omega +\tau_\phi^{-1}| \simeq D/A$,
we expect a smooth cross-over between these two limiting cases.

In the limiting cases one is able to obtain explicit
results for the averaged weak localization correction $\delta D$ in the following way.
Depending
on whether $A|-i\omega +\tau_\phi^{-1}|/D$ is smaller or larger than unity, it is
convenient to replace the zero-field cooperon ${\cal C}^{(0)}_\omega$
in the solution for ${\cal C}^{(2)}_\omega$, 
Eq.\ (\ref{c2int}), by its
diffusion mode representation (\ref{czero}) or its Fourier
representation for an infinite 2D system
\begin{equation}\label{coopinf}
{\cal C}^{(0)}_\omega( \mbox{\boldmath $r$}, \mbox{\boldmath $r$}')=
\int \frac{d^2\mbox{\boldmath $q$}}{(2\pi)^2} \,
\frac{e^{i\mbox{\boldmath $q$}(\mbox{\boldmath $r$}-\mbox{\boldmath $r$}')}}{-i\omega
+Dq^2+\tau_\phi^{-1}}  ,
\end{equation}%
respectively.
In the former case,
we are able to separate the large contributions due to the zero-mode
from the corrections resulting from all other modes.
In the latter, the particle diffuses on scales small compared to
the size of the system; hence, the boundary conditions imposed
on a finite dot can be disregarded.

\subsubsection{Small dot case}

Substituting Eq.\ (\ref{czero}) in ${\cal C}^{(2)}_\omega$, Eq.\ (\ref{c2int}), yields
\begin{eqnarray}\label{caver}
\int d^2 \mbox{\boldmath $r$} \, 
{\cal C}_\omega( \mbox{\boldmath $r$}, \mbox{\boldmath $r$}) &= &
\sum\limits_{m=0}^{\lambda_m \lesssim l^{-2}}
\Biggl[ -i\omega+ D\lambda_m + \tau_\phi^{-1}+ \\
&& \left. + D \int d^2 \mbox{\boldmath $r$} \, 
\psi_m^2(\mbox{\boldmath $r$})(2e\mbox{\boldmath $A$}(\mbox{\boldmath $r$}))^2 +
D^2 \sum\limits_{n=0}^{\lambda_n \lesssim l^{-2}}
\frac{(F_{nm}-F_{mn})^2}{-i\omega+ D\lambda_n + \tau_\phi^{-1}}
\right]^{-1} , \nonumber
\end{eqnarray}%
where
\begin{equation}\label{vmn}
F_{mn}=
\int d^2 \mbox{\boldmath $r$} \, \psi_m (\mbox{\boldmath $r$}) 
i2e\mbox{\boldmath $A$}(\mbox{\boldmath $r$}) \nabla \psi_n (\mbox{\boldmath $r$})  .
\end{equation}%
For $A | -i \omega +\tau_\phi^{-1}|/D <1$, we restrict the sum to the $m=0$ term 
and obtain
\begin{equation}\label{smalldot}
\delta D  =
- \frac{D}{\pi\hbar \nu A} \, \frac{1}{-i \omega +\tau_\phi^{-1} + c_1 DA/ l_B^4 }  ,
\end{equation}%
where $c_1$ is a real positive constant of order unity.
The lowest eigenmode $\psi_0$
leads to an increase
of the weak localization corrections as the
area $A$ of the dot decreases. This
behavior results from the fact that the boundaries of an
isolated system cause no phase-breaking.
In the case where the dot is connected
to leads, the summation over modes starts at $m=1$ and
yields \cite{Abrahams79} $\delta D=-(\pi\hbar\nu )^{-1}\ln{(A/l^2)}$  for $B=0$.
That is, the correction term decreases as the dot shrinks.
The reduction of the size of the system is then
associated with contacts approaching each other, which leads in turn
to a decrease of the phase coherence.
The scales of the critical magnetic field
at which the phase coherence is significantly reduced are different
for isolated and open systems as well.
The critical magnetic field can be deduced from Eq.\ (\ref{caver})
by equating the two magnetic field dependent terms
[which are of the same order]
to the first one.
For an isolated dot, the first term is 
of order $|-i\omega+\tau_\phi^{-1}|$, Eq.\ (\ref{smalldot}),
whereas it is given by
$D/A$ for a dot coupled to leads. This  results in the estimates
\begin{equation}\label{critmag}
(l_B^4)^{iso} \simeq A \, {\rm min}\{ D/\omega, l_\phi^2 \}
\qquad {\rm and} \qquad
(l_B^4)^{lead} \simeq A^2  .
\end{equation}%
The critical magnetic field for an isolated small dot is thus much larger than
for a dot with leads.

\subsubsection{Big dot case}

For $A|-i\omega+\tau_\phi^{-1}|/D>1$, we substitute
Eq.\ (\ref{coopinf}) in ${\cal C}^{(2)}_\omega$, Eq.\ (\ref{c2int}),
and obtain
\begin{equation}\label{bigdot}
\delta D =
\frac{c_2}{\pi \hbar\nu } \,
\ln{\left\{ \tau [ -i \omega +\tau_\phi^{-1} + c_3 DA/ l_B^4 ] \right\} }  ,
\end{equation}%
where $c_2$ and $c_3$ are real positive constants of order unity. In this case,
the zeroth mode plays no role; Eq.\ (\ref{bigdot}) is valid independently
of the boundary conditions imposed.

The results (\ref{smalldot}) and (\ref{bigdot}) for $\delta D$
can be substituted in Eqs.\ (\ref{etaabinf}), (\ref{etaabqa}),
and (\ref{etascqa}) for the absorption coefficient and the cross-sections,
respectively, in order to account explicitly for the weak localization corrections.
Note, however, that the characterization as ``big" or
``small" dot does not necessarily apply simultaneously to both the diffusion
propagator [Eq.\ (\ref{difpro2})] and the cooperon [Eq.\ (\ref{czero})].
In particular, for $\omega \tau_\phi \ll 1$, there exists the situation
where the
diffusion propagator [responsible for the classical contributions]
is determined by the zeroth mode since $A\omega/D \ll 1$, whereas the
cooperon [determining $\delta D$]
behaves bulk-like since $A\tau_\phi^{-1}/D \gg 1$.

\section{Numerical calculations}\label{numerics}

To compute numerically
the scattering and the absorption cross-sections, we have used an accurate
inversion of the dielectric function $\epsilon(\mbox{\boldmath $r$},\mbox{\boldmath $r'$})$, 
which in turn yields
the total potential $V$, see Eqs.\ (\ref{vinvers}) and
(\ref{epsin}).
This exact procedure covers small and large values of the
parameter $qL$. In this sense, the numerical results bridge the gap
left by the analytical study of limiting cases in Sec.\ \ref{cross}, and
provide information about the angular dependence
of the cross-sections.
The scaling of the numerically calculated cross-sections with
respect to $A,q,D$ etc.\ can be used to confirm the predictions derived analytically,
cf.\ Eqs.\ (\ref{etaabinf})--(\ref{etascqa}).
In Figs.\ \ref{logfig}--\ref{scfig}
discussed below, results for different wave vectors $q$
are shown; all other parameters are kept fixed.
This means that the two dimensionless quantities $qL$ and
$A\omega/D$ vary. Since the latter, even for the largest wave vectors
used is smaller than unity, 
it is essentially the effects of the variation of $qL$ which
we focus on. In this small dot case, the weak
localization corrections are essentially described by the quantity
$1-\Re[\delta D]/D$, cf.\ Eqs.\ (\ref{etaabqa}) and (\ref{etascqa}).
The dependence of this quantity on magnetic field, frequency and temperature
is discussed in connection with Fig.\ \ref{ddfig}.

The dot has the shape of a square (lengths $L_x=L_y$)
or a rectangle ($L_x > L_y$).
In the first case emphasis is put 
on the 
angular dependence associated with the value of the parameter $qL$.
The eigenfunctions are
given by $\psi_{m_x,m_y}(x,y)\sim \cos{(m_x \pi x/L_x)} \cos{(m_y \pi y/L_y)}$,
where $m_x$ and $m_y$ are positive integers starting from zero.
The number of diffusion modes for each direction is restricted to, e.g.,
$m_x^{max}=m_y^{max}=12$ for the square and to
$m_x^{max}=20$, $m_y^{max}=8$ for the rectangle discussed in Fig.\ \ref{abfig}. 
Hence, the total number of different modes,
which coincides with the dimension of the matrices $\epsilon_{mn}, G_{mn}$, etc.,
is $12^2=144$ or $20 \times 8=160$, respectively.

The numerical calculations are based on values for
the physical quantities which apply to GaAs/Al${}_x$Ga${}_{1-x}$As
heterostructures:
$\nu=1.55 \times 10^{10}$ meV${}^{-1}$cm${}^{-2}$,
$s=2.7\times 10^{5}$ cms${}^{-1}$, $D=140$ cm${}^2$s${}^{-1}$ [$l=0.1$ $\mu$m], 
$\tau=0.4 $ ps,
$\tau_\phi=0.03$ ns, $d=0.1$ $\mu$m, $a_B=10.6$ nm, $\epsilon_\circ=12.8$, $\epsilon_1=1$,
and, e.g., $L_x=L_y=0.66$ $\mu$m for the square dot.
The phase coherence time (corresponding to $T=0.1$ K and
$l_\phi=0.63$ $\mu$m)
and the diffusion coefficient
for a low mobility electron gas are taken from Ref.\ \cite{Beenakker91}.
The selection of a small diffusion coefficient is dictated by the 
condition that the size of the dot is larger than the mean free path,
but small enough to allow for $qL< 1$ and $qL>1$ in the
range of reasonable sound frequencies.
For these frequencies and the quantities given above, $\omega$ and 
$\tau_\phi^{-1}$ are smaller than $D/A$. 
Thus, the diffusion propagator [Eq.\ (\ref{difpro2})]
and the cooperon [Eq.\ (\ref{czero})] are essentially determined by the zeroth mode.
Since the very existence of the zeroth mode
relies on the assumption of an isolated dot, it is this physical regime which
puts most emphasis on the noninvasive character of the proposed SAW
measurement. 
Posing the requirement $q<l^{-1}$ yields $q\approx 10^5$ cm${}^{-1}$
($\omega \approx 2\pi \times 4.3$ GHz) as the upper limit 
for the applicability of the diffusion
approximation. A lower limit $\omega > \Delta$ could arise from the
finiteness of the mean level spacing $\Delta=(\nu A)^{-1}$
in the quantum dot. For $A\simeq 1$ $\mu$m${}^2$, we have $\Delta \simeq 6$ $\mu$eV
corresponding to $\omega \simeq 10^{10}$ s${}^{-1}$.
We argue, however, that inelastic level broadening smears out
the discreteness of the one-particle levels, rendering the spacing
$\Delta$ irrelevant. Indeed,  using the phase coherence time
introduced above,
we find
$\hbar /\tau_\phi \simeq 20$ $\mu{\rm eV }> \Delta$. That is,
a lower limit for the frequency is not required.

Figure \ref{logfig} shows in a double-logarithmic plot the absorption
and the scattering cross-sections, Eqs.\ (\ref{etaab}) and (\ref{etasc}),
as a function of the wave vector. Here and in Figs.\ \ref{abfig}
and \ref{scfig} the term $\hat{q}_x\hat{q}_y$
of the electron-surface phonon vertex, Eq.\ (\ref{gamma}), has been
replaced by its maximum $\case1/2$; its angular dependence is not
taken into account because it depends on the orientation of the dot
with respect to the lattice axes. Quantum corrections for $B=0$
are incorporated and $L_x=1.7$ $\mu$m and $L_y=1$ $\mu$m.
The inset in Fig.\ \ref{logfig} shows the correct aspect ratio
of the dot, the direction of the incoming SAW (unlabeled arrow),
and the direction
of the outgoing SAW for a scattering angle of 30${}^\circ$ (arrow 2).
The corresponding absorption cross-section $\eta_{abs}(\mbox{\boldmath $q$})$
is given by curve 1. The part to the left of label 1
exhibits the behavior $\eta_{abs} \sim q^3$, confirming the power law
predicted by Eq.\ (\ref{etaabqa}) for the limit $qL \ll 1$.
The remaining part of the curve corresponds to $\eta_{abs} \sim q$.
This behavior is anticipated in the regime $qL \gg 1$;
see Eq.\ (\ref{etaabinf}). The oscillations of $\eta_{abs}(\mbox{\boldmath $q$})$
represent geometric resonances. Maxima appear around $qL_x=2\pi m$,
$m=1,2,\ldots$, whereas minima occur for $qL_x=\pi(2m+1)$. The
curve labeled 2 (corresponding to the direction 2)
represents results for the scattering cross-section.
It follows the power law $\eta_{sc} \sim q^5$, Eq.\ (\ref{etascqa}),
in the limit $qL \ll 1$. For $qL  \gg 1$,
the scattering cross-section
behaves in a sample and angle specific way. 
In this case, the magnitude of $\eta_{sc}(\mbox{\boldmath $q$}'\ne\mbox{\boldmath $q$})$ is small
compared to the forward scattering component which continues to grow as $q^5$,
cf.\ relation (\ref{etascinf}) and Fig.\ \ref{scfig}.
Taking the results for both $\eta_{abs}$ and $\eta_{sc}$ into account,
one may conclude that the limiting equations for 
$qL \ll 1$ and $qL  \gg 1$ given in Sec.\ \ref{cross}
represent a good description up (or down) to $qL\approx 2\pi$.

Figure \ref{abfig} shows results for the absorption cross-section, Eq.\ (\ref{etaab}),
of a rectangular (main plot) and a square dot (inset). The size of the dots
is given by $L_x=1.2$ $\mu$m and $L_y=0.4$ $\mu$m and $L_x=L_y=0.66$ $\mu$m, respectively.
The abscissa of the figure agrees with the $x$-axis of the dot.
Figure \ref{abfig} represents a polar diagram for $\eta_{abs}(\mbox{\boldmath $q$})$, i.e.\
the distance of a data point from the origin corresponds to
the magnitude of $\eta_{abs}$ [given in $\mu$m],
whereas the orientation of the wave vector
$\mbox{\boldmath $q$}$ of the SAW agrees with the direction of the line joining
the data point and the origin. The curves labeled 1, 2, and 3 correspond
to the wave vectors $q_1=10^4$ cm${}^{-1}$, $q_2=5\times 10^4$ cm${}^{-1}$, and
$q_3=10^5$ cm${}^{-1}$, respectively. The magnitude of the curves labeled 1
is increased by a factor of 20. The main plot in Fig.\ \ref{abfig}
for the rectangular dot shows a
strong anisotropy of the absorption cross-section
for the two smaller wave vectors. For instance,
$\eta_{abs}(q_1\hat{x})$ and $\eta_{abs}(q_1\hat{y})$ deviate by
a factor of about 20 from each other. This pronounced anisotropy
can be attributed to the fact that the relevant size of the
dot (either $L_x$ or $L_y$) enters Eq.\
(\ref{etaabqa}) for $\eta_{abs}$ as a high power in the limit $qL < 1$.
In contrary, for the largest wave vector $q_3$, $\eta_{abs}(\mbox{\boldmath $q$})$
exhibits only a minor dependence
on the angle of incidence, reflecting that only the total area is relevant
in the regime $qL  \gg 1$, cf.\ Eq.\ (\ref{etaabinfp}). That is, for
an approximate isotropy of $\eta_{abs}(\mbox{\boldmath $q$})$ to appear,
both $qL_x$ and $qL_y$ have to be sufficiently larger than unity.
As can be seen from the inset of Fig.\ \ref{abfig}, the situation
is somewhat different in the case of a square dot. Here, $\eta_{abs}(\mbox{\boldmath $q$})$
is completely independent of the direction of incidence for
the smallest wave vector $q_1$. The absorption cross-section
acquires a weak angular dependence with increasing wave vector which
can be attributed to geometrical resonances occurring for
$qL \ge 2\pi$.

All the curves presented in Fig.\ \ref{abfig} include weak localization
corrections to the density-density correlator, Eq.\ (\ref{bewl}), for zero magnetic
field. As discussed in Sec.\ \ref{cross}, the absorption is increased by
these effects. This enhancement is expected to be given
by a factor $1-\Re[\delta D]/D$ independent of $q$ for the
``small-dot" case, cf.\ Eqs.\ (\ref{etaabqa}) and (\ref{smalldot}).
Indeed, all numerical calculations indicated that the curves
with and without quantum corrections exhibit practically 
the same angular
dependence, merely the magnitudes differ by a constant factor.
Because of this, we do not incorporate in Fig.\ \ref{abfig} curves
which deviate from each other only with respect to
the magnitude of the weak localization corrections. Instead,
we discuss the enhancement factor $1-\Re[\delta D]/D$ separately
below.

Figure \ref{scfig} presents results for the elastic scattering cross-section, Eq.\
(\ref{etasc}), of a square dot. $\eta_{sc}(\mbox{\boldmath $q'$},\mbox{\boldmath $q$})$
is given in $\mu$m. A
polar representation is chosen with respect to the direction of 
the outgoing surface phonon $\hat{\mbox{\boldmath $q'$}}$.
The direction of incidence, $\hat{\mbox{\boldmath $q$}}$, is fixed
and  is oriented perpendicularly to one of the edges  of the square dot,
see the inset. The labels 1, 2, and 3 of the curves indicate the
wave vectors used,
$q_3=2q_2=10q_1=10^5$ cm${}^{-1}$. However, due to the significant
dependence of $\eta_{sc}$ on the magnitude of $q$, a
normalization of the curves different from that of 
Fig.\ \ref{abfig} is required: the data of curve 1 are multiplied
by $10^3$, while that of curve 3 are divided by 15 and those of curve 2 are not
changed.
Generally, the scattering cross-section of the square dot shows a weak dependence
on the angle of incidence but varies considerably with the
scattering angle $\theta$, i.e.\ 
$\theta= < \!\!\! ) \,\, (\hat{\mbox{\boldmath $q$}}',\hat{\mbox{\boldmath $q$}})$.
In particular, the angular dependence of $\eta_{sc}$ in the regime
$qL < 1$ [curve 1] is given by $\eta_{sc} \sim \cos^2{\theta}=
(\hat{\mbox{\boldmath $q$}}\hat{\mbox{\boldmath $q$}'})^2$.
This behavior 
could be anticipated from 
the equations given in Sec.\ \ref{cross}, taking into account 
not only the magnitude of the
wave and spatial vectors but also their orientation.
With increasing wave number $q$, the $\cos^2{\theta}$-law
is gradually replaced by an enhancement of forward scattering and
a suppression of back scattering. [This is known in the
theory of electromagnetic fields as Mie effect, cf.\
Ref.\ \cite{Born75}, p.\ 654.]
For $q=q_2$, only a small
back-scattering component is left. For even larger $q$, e.g.\ 
$q=q_3$, this component
is not resolved on the scale of Fig.\ \ref{scfig}. 
This confirms Eq.\ (\ref{etascinf}).
In agreement with our qualitative analysis, quantum corrections are
extremely small for the parameters introduced above. The curves in Fig.\ \ref{scfig}
correspond therefore essentially to the classical part of the
scattering cross-section.

In the small dot case
under consideration, the weak localization corrections
to the absorption cross-section are significant, whereas
the scattering cross-section remains practically unaffected.
According to the analytically derived expressions
(\ref{etaabinf}) and (\ref{etaabqa}) and the numerical calculations, these
corrections to $\eta_{abs}$ are well described by the enhancement
factor $1-\Re[\delta D]/D$. To illustrate the dependence
of the weak localization corrections to $\eta_{abs}$
on the perpendicular magnetic field $B$, the frequency
and the temperature,
we have evaluated $1-\Re[\delta D]/D$ for
the cooperon expression (\ref{caver}).
For the small dot, $A|-i\omega +\tau_\phi^{-1}|/D \ll 1$,
we expect $\delta D$ to be determined by the lowest mode.
According to Eq.\ (\ref{smalldot}),
\begin{equation}\label{smalldotsp}
-\delta D/D= \Delta \tau_\phi/\pi \hbar =0.23 ,
\end{equation}%
where we have used $\omega < \tau_\phi^{-1}$ and $B=0$.
The right-hand side follows from the numerical values introduced above.
In Fig.\ \ref{ddfig}, $1-\Re[\delta D]/D$ is shown as
a function of the magnetic field.
The dot is assumed to be of a square shape with $L_x=L_y=0.66$ $\mu$m.
All other parameters are as defined above.
The deviations of the numerical results
from the estimate (\ref{smalldotsp}) arise from the contributions of the
higher modes.
The assumption $l_B>L$, used to derive the terms in the cooperon
which depend on the magnetic field, is valid for $B<1.4 $ mT.
The continuation of the curves to stronger magnetic fields
can only serve as an indication for the further suppression
of the weak localization corrections with increasing $B$.
The three curves in Fig.\ \ref{ddfig} show how the quantity
$1-\Re[\delta D]/D$ decreases with increasing frequency.
Assuming \cite{Beenakker91} $\tau_\phi \sim T^{-1}$, 
the increase of the temperature from $T=0.1$ K to $T=1$ K corresponds
to a reduction of the phase coherence length
from $l_\phi=0.63$ $\mu$m [which is used in Fig.\ \ref{ddfig}] to
$l_\phi=0.2$ $\mu$m. The latter value is significantly smaller
than the size of the dot, $L=0.66$ $\mu$m, leading to 
$1-\Re[\delta D]/D\approx 1.07$ for $B=0$. The relative differences 
between the three curves shown in Fig.\ \ref{ddfig} become much smaller as well.

\section{Conclusions}
We have calculated the absorption and the 
scattering cross-sections, $\eta_{abs}(\mbox{\boldmath $q$})$ 
and $\eta_{sc}(\mbox{\boldmath $q$}',\mbox{\boldmath $q$})$, respectively,
of a surface acoustic wave [SAW] for an isolated quantum dot. 
The dependence
of these quantities on weak localization corrections
has been found. In addition, we have calculated the weak localization corrections
to the attenuation coefficient $\Gamma_{\mbox{\boldmath $q$}}$ of an extended 2DEG
[Eq.\ (\ref{etaabinf})].
Since these corrections can (at least) approximately be expressed in terms of $\delta D$,
the spatial average of the corresponding change of the diffusion coefficient
[Eq.\ (\ref{deltad})],
they are given in a similar
way as those to the conductivity. One can therefore use results derived
in that case to establish easily the dependence of the 
cross-sections and $\Gamma_{\mbox{\boldmath $q$}}$
on spin-orbit scattering, scattering by magnetic impurities, etc. \cite{Altshuler92}
We emphasize the weak localization corrections because they are expected
to play a significant role in the experimental investigation of the 
effects discussed in this paper. Indeed, though the classical
cross-sections depend strongly on $A$ and $\omega$,
these parameters are fixed once the dot and interdigital transducers
are defined on a sample. Measurements at different frequencies or at other
sizes of the dot require the preparation of different samples. 
The proximity technique \cite{Schenstrom88,Wixforth89}
may reduce the experimental effort, but it still
provides only a set of discrete frequencies $\omega$ at which a certain
sample can be studied. Conversely, it poses no serious
problems to vary continuously the temperature and the magnetic field
which both affect only the weak localization corrections.
Note also that the SAW technique allows precise measurements
of the relative changes of the transmitted wave intensity, whereas the
absolute attenuation is much less easily detectable. That is,
the large but constant classical effects are generally
more difficult to resolve than the quantum corrections
which can be ``tuned" by external parameters.
For example, the measurement of the absorption cross-section (or $\Gamma_{\mbox{\boldmath $q$}}$) 
as a function of the temperature yields directly the dependence
of the phase coherence time on the temperature. For typical
experimental values, $\omega \tau_\phi < 1$, and, hence, the parameter
$A/\l_\phi^2$ determines whether the dot
has to be considered as a small or big one.
Consequently, the temperature can also shift the dot from one regime
to another.

Depending on the sensitivity of the methods used
to measure SAW's, an experimental investigation
might be carried out for an array of quantum dots rather than a single dot.
Since the electron-phonon coupling is weak (even for the piezoelectric interaction),
it is reasonable to assume that the response of a dot array to a SAW
can be represented by a superposition of the effects associated with
isolated dots. To underscore this point, let us give some numerical
estimates for the SAW attenuation and the electron heating. 
To estimate whether the calculated cross-sections are within
the experimental sensitivity,
we convert the absorption cross-section
to an attenuation coefficient by $\Gamma_{\mbox{\boldmath $q$}} \simeq \eta_{abs}/A$, cf.\
Eq.\ (\ref{etaabinfp}). This amounts to covering densely
the area between the transducers with quantum dots. Using 
$\eta_{abs} \simeq 10^{-4}$ $\mu$m  and $A\simeq 1$ $\mu$m${}^2$ yields an attenuation
of about 10 dB/cm. The relative change of the attenuation due to
weak localization effects is then about 1 dB/cm.
This value is about 10 times larger than the highest
resolution achieved, suggesting that the signal of a much less dense arrangement
of dots can be measured.

To estimate the effect of electron heating,
one has to compare the temperature of the dot with 
$\Delta T\equiv I\eta_{abs} \tau_{\epsilon}/k_B$, where $I$
is the flux intensity of the incoming surface wave,
$\tau_{\epsilon}$ is the energy relaxation time
and $k_B$ is the Boltzmann constant.
Using $w \approx 2$ mm for the length and the width of a macroscopic
SAW delay line and $P \approx 1$ $\mu$W for the total SAW intensity, we determine
$I=P/w$ [the experimental values are taken from
Ref.\ \cite{Schenstrom88}]. The energy relaxation time $\tau_\epsilon$
can roughly be identified with the phase coherence time $\tau_\phi$.
[We note that this is a good estimate in the case where both $\tau_\epsilon$
and $\tau_\phi$ result from electron-electron scattering; see
Ref.\ \cite{Altshuler92}.] Using the values $\tau_\phi=30$ ps and 
$\eta_{abs} \simeq 10^{-4}$ $\mu$m, we obtain $\Delta T\simeq 0.1$ K which represents
a significant change in the temperature range of interest.

\section*{Acknowledgments}

Financial support by the German-Israeli Foundation, the
Fund for Basic Research administered by the Israel Academy
of Sciences and Humanities, and the Deutsche Forschungsgemeinschaft (A.~K.)
is gratefully acknowledged.
We thank
Y.\ Galperin, A.\ Kamenev, D.\ Khmelnitzkii, D.\ Klakow, C.\ Rocke, M.\
Rotter, A.\ Tilke, and A.\ Wixforth for valuable discussions.

\newcommand{\noopsort}[1]{} \newcommand{\printfirst}[2]{#1}
  \newcommand{\singleletter}[1]{#1} \newcommand{\switchargs}[2]{#2#1}

\begin{figure}
\caption{\label{ephver}
Electron--surface-acoustic phonon interaction processes.
The phonons are represented by wavy lines, and the electrons
are depicted by straight lines. 
Diagram (a) shows the absorption of an incoming phonon 
(wave vector \protect\boldmath{$q$}, energy
$\protect\omega$).
Diagrams (b) and (c) show scattering processes with two different 
intermediate states.
}
\end{figure}
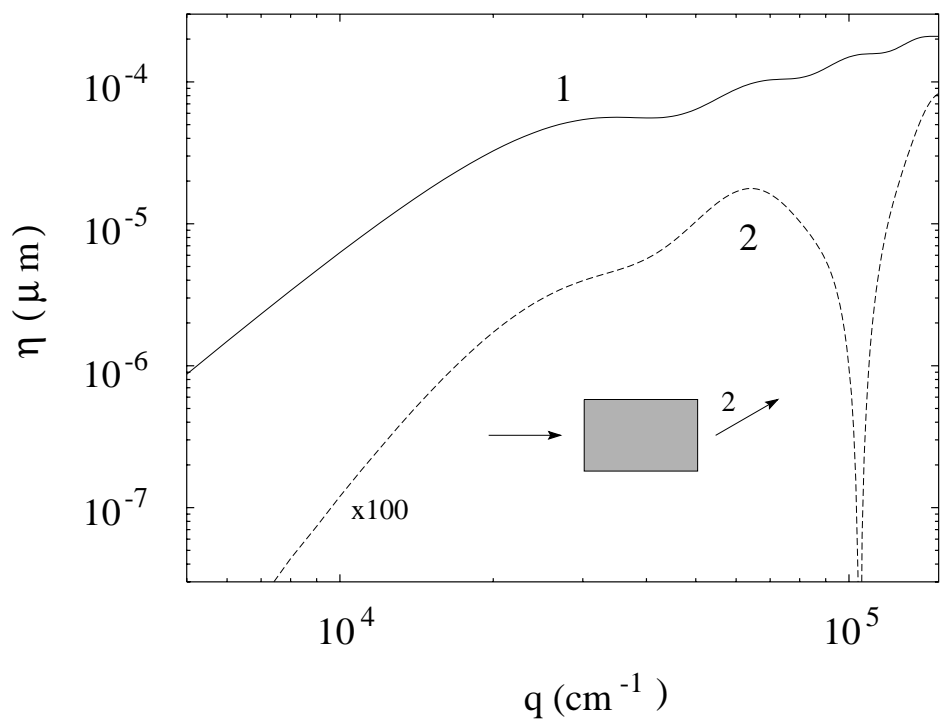
\begin{figure}
\caption{\label{logfig}
Double logarithmic plot of the absorption and the scattering cross-sections
as a function of the wave vector.
}
\end{figure}
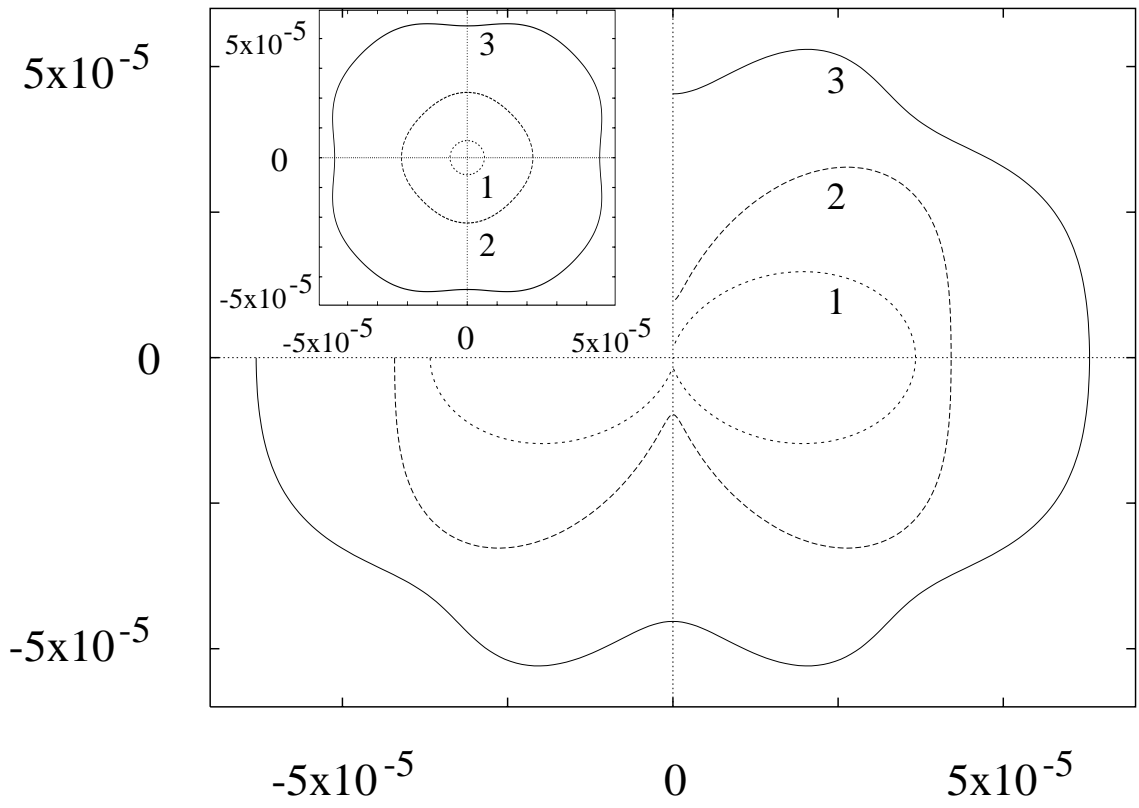
\begin{figure}
\caption{\label{abfig}
Polar diagrams of the absorption cross-section 
$\protect\eta\protect_{abs}($\protect\boldmath{$q$}$)$,
Eq.\ (\protect\ref{etaab}), of a rectangular (main plot) and a
square dot (inset). $\protect\eta\protect_{abs}($\protect\boldmath{$q$}$)$ 
[in $\mu$m] is
a function of the angle of incidence of the SAW and is given for three
different wave numbers q.
}
\end{figure}
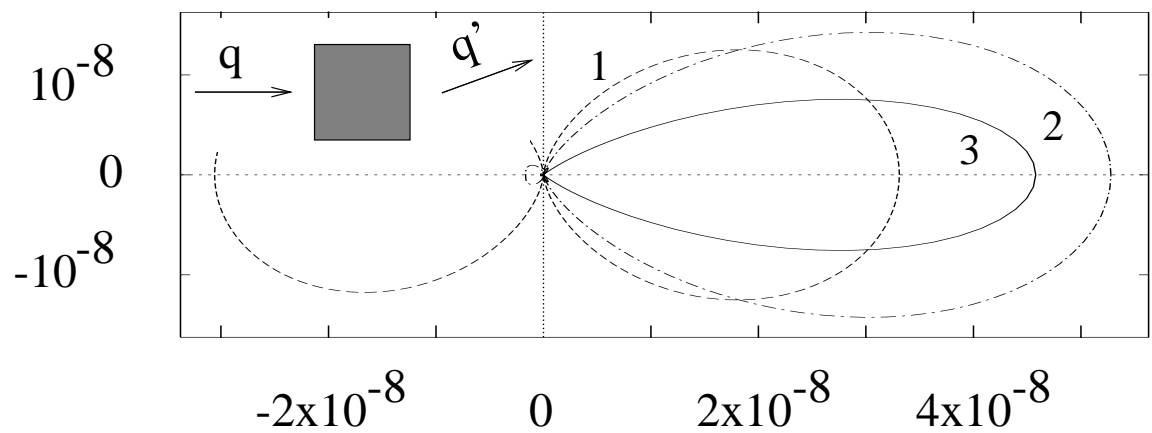
\begin{figure}
\caption{\label{scfig}
Polar diagram of the elastic scattering cross-section 
$\protect\eta\protect_{sc}($\protect\boldmath{$q$}$',$\protect\boldmath{$q$}$)$,
Eq.\ (\protect\ref{etasc}), of a square dot. $\protect\eta\protect_{sc}$
[in $\mu$m] is considered as a function of the direction \protect\boldmath{$q'$} of
the outgoing wave; the angle of incidence (\protect\boldmath{$q$}) is kept fixed,
see inset.
}
\end{figure}
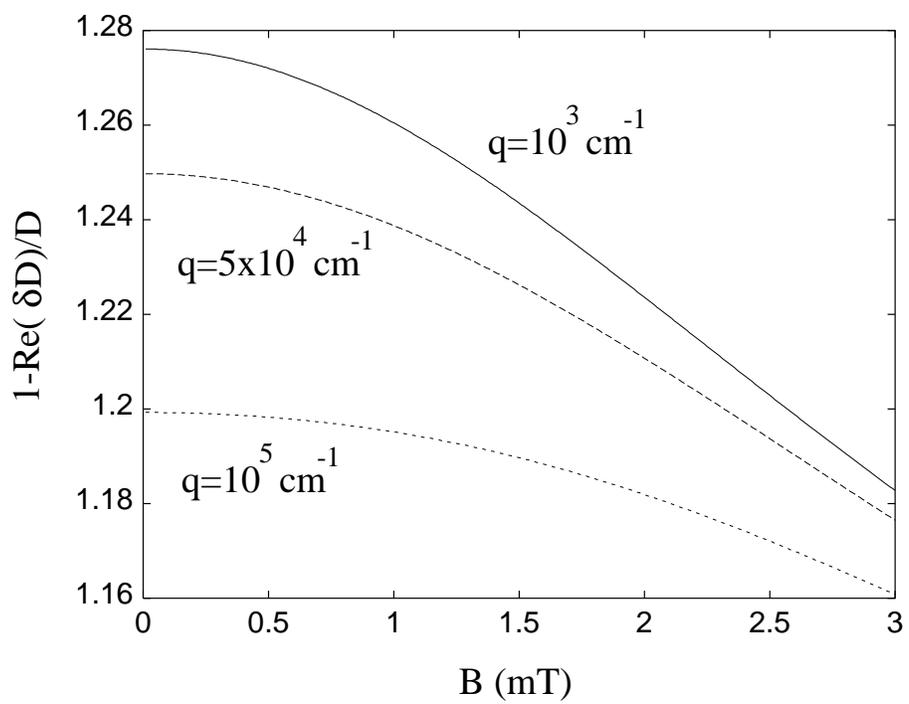
\begin{figure}
\caption{\label{ddfig}
Weak localization enhancement $1-\protect\Re [\protect\delta D]/D$,
Eq.\ (\protect\ref{deltad}), of
the absorption cross-section, Eq.\ (\protect\ref{etaabqa}), as a function
of the magnetic field $B$ for three different wave numbers.
}
\end{figure}

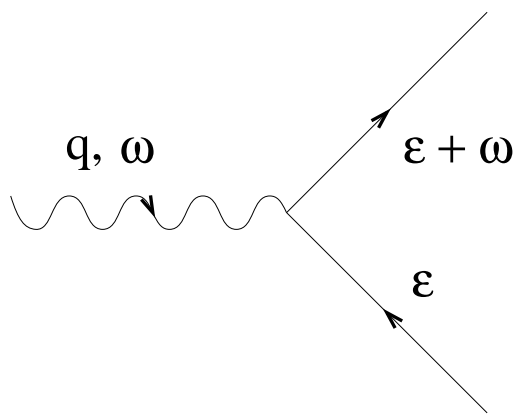
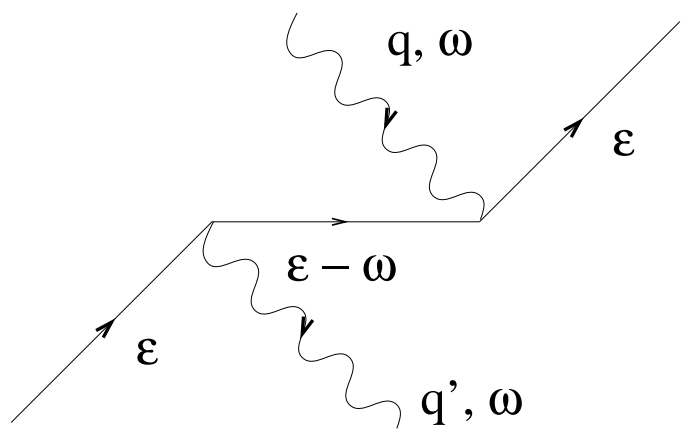
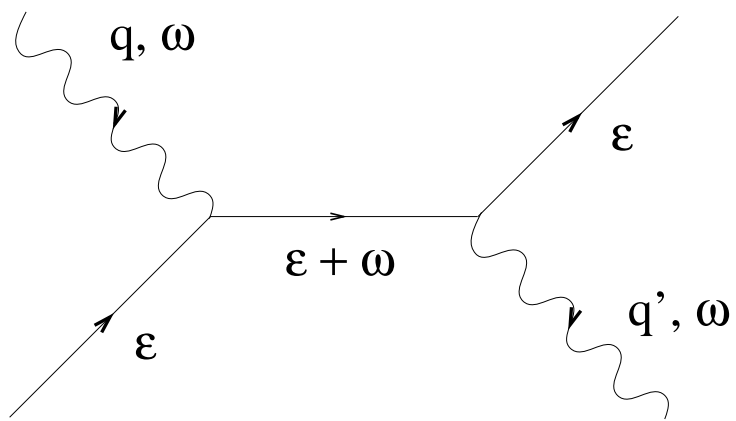
\end{document}